\documentclass[draft]{agujournal2018}
\usepackage{apacite}
\usepackage{url} 
\usepackage{lineno}
\draftfalse

\journalname{JGR: Space Physics}

\begin{document}

%
%


\title{Magnetic Reconnection  within the Boundary Layer of a Magnetic Cloud in the Solar Wind }

%
%




\authors{Zolt\'an V\"{o}r\"{o}s\affil{1,2}, Ali Varsani\affil{1}, Emiliya Yordanova\affil{3}, Yury L. Sasunov\affil{1,4,5}, Owen W. Roberts\affil{1}, \'Arp\'ad Kis\affil{2}, Rumi Nakamura\affil{1}, Yasuhito Narita\affil{1} }

\affiliation{1}{Space Research Institute, Austrian Academy of Sciences, Graz, Austria}
\affiliation{2}{Institute of Earth Physics and Space Science, E\"otv\"os Lor\'and Research Network, Sopron, Hungary}
\affiliation{3}{Swedish Institute of Space Physics, Uppsala, Sweden}
\affiliation{4}{Department of Astrophysics, University of Vienna, Austria}
\affiliation{5}{Skobeltsyn Institute of Nuclear Physics, Lomonosov Moscow State University, Russian Federation}





\correspondingauthor{Zolt\'an V\"or\"os}{zoltan.voeroes@oeaw.ac.at}




\begin{keypoints}
\item Magnetic reconnection
\item Magnetic cloud
\end{keypoints}

%
%


\begin{abstract}

The twisted local magnetic field at the front or rear regions of the magnetic clouds (MCs) associated with interplanetary coronal mass ejections (ICMEs) is often nearly opposite to the direction of the ambient interplanetary magnetic field (IMF). There is also observational evidence for magnetic reconnection (MR) outflows  occurring  within the boundary layers of MCs. In this paper a MR event located at the western flank of the MC occurring on 2000-10-03 is studied in detail. Both the large-scale geometry of the helical MC and the MR outflow structure are scrutinized in a detailed multi-point study. The ICME sheath is  of hybrid propagation-expansion type.  Here the freshly reconnected open field lines are expected to slip slowly over the MC resulting in plasma mixing at the same time. As for MR, the current sheet geometry and the vertical motion of the outflow channel between ACE-Geotail-WIND spacecraft was carefully studied and tested. The main findings on MR  include: (1) First-time observation of non-Petschek-type slow-shock-like  discontinuities in the inflow regions; (2) Observation of turbulent Hall magnetic field associated with a Lorentz force deflected electron jet; (3) Acceleration of protons by reconnection electric field and their back-scatter from the slow shock-like discontinuity; (4) Observation of relativistic electron near the MC inflow boundary/separatrix; these electron populations can presumably appear as a result of non-adiabatic acceleration, gradient B drift and via acceleration in the electrostatic potential well associated with the Hall current system; (5) Observation of Doppler shifted ion-acoustic and Langmuir waves in the MC inflow region.
\end{abstract}

%
%

 \section{Introduction}

Coronal Mass Ejections (CMEs) originate in energetic eruptions associated with solar flares, filaments or prominences in the Sun's lower corona.  Before the eruption the coronal magnetic field is stressed and highly twisted. After the eruption the magnetic field configuration, possibly via magnetic reconnection (MR),  relaxes to a less tense state. Eventually, part of the  relatively colder and denser plasma previously confined in magnetic field falls back onto the solar surface.
During the eruption electromagnetic energy is converted into kinetic energy of plasma and particles are accelerated to high energies. The source active regions and the fast outward propagating eruptive plasmas containing the often twisted magnetic field have been observed remotely for more than 30 years in white and extreme ultraviolet light, X-rays, radio emissions, through vector magnetograms and by using the Michelson-Doppler Imager \citep[e.g.][]{schwenn06}. The remote measurements indicate that the generic features of fast CMEs are relatively well understood. CMEs are structured having a shock and compressed plasma at their leading edge, followed by low density often helical magnetic field and dense gas at their trailing edge. The structure, evolution and propagation properties of  CMEs obtained from remote sensing observations and models have been thoroughly described in several reviews and numerous papers referenced therein \citep{chen11, webb12, kilpua17a, manchester17}.

Although the details of the  outward propagation of CMEs  are not fully understood, the ejecta are observed through in-situ measurements at larger heliocentric distances including the near-Earth space as well. According to these observations, the Interplanetary Coronal Mass Ejections (ICMEs) are also structured, usually having a shock wave at their front edge followed by a compressed and turbulent sheath region and a helical flux rope or Magnetic Cloud (MC). Since fast ICMEs with embedded helical MCs are particularly geoeffective there is a strong interest in understanding and predicting the propagation of these transients in space weather context as well \citep[e.g.][]{marubashi00, kilpua17b, mostl18, vourlidas19}.

The local in-situ field, plasma, energetic particle, charged state and compositional signatures of ICMEs differ significantly from event to event \citep{zurbuchen06}. CMEs propagating faster than the fast magnetosonic speed relative to the ambient solar wind generate fast-mode forward shocks, accelerating particles by various shock-related  acceleration mechanisms \citep{zank00}. Downstream of a shock and ahead of an MC, the sheath region contains compressed turbulent plasma,  magnetic field fluctuations, wavy and coherent structures. The occurrence and magnitude of these fluctuations and intermittent structures depend on the geometry of and on the distance from the shock and on the speed of ICME relative to the ambient solar wind speed in a complex manner which is not fully understood yet \citep{liu06a, kilpua20}.

Contrarily to the sheath regions, MCs contain large-scale rotations but reduced level of magnetic field fluctuations.
Other generic signatures of magnetic clouds are the increased magnetic field, decreased ion density and temperature, low plasma beta, bi-directional suprathermal electrons and $\sim$MeV ions, enhanced $He^{2+}/H^{+}$ ratio, elevated oxygen charge states $O^{7+}/O^{6+}$, etc. \citep{zurbuchen06}. One possibility to explain the presence of the bi-directional suprathermal electrons inside MCs is that the spacecraft  crosses  closed field lines which are rooted at the Sun at both ends. This suggests that the local signatures can be associated with remote processes including those in the solar atmosphere. Also, the in-situ ionic charge states or the elemental abundances can be determined by the details of the generation of CMEs on the Sun or by elemental composition in the pre-CME solar atmosphere, respectively \citep{wimmer06}. However, because of the crossing geometry or the complexity of interactions, not each of the above outlined in-situ signatures of  MCs are  always observed \citep{marubashi00, burlaga02}.

The large-scale interactions and reorganisations of the magnetic field often lead to local physical conditions which trigger MR. Although the first direct observations of MR were accomplished at the Earth´s magnetopause by \citet{paschmann79} and \citet{sonnerup81} four decades ago,
the local signatures of MR in the solar wind were overlooked, or at best, MR was not considered energetically important  until 2005 \citep{gosling05, gosling12}. Rare multi-point observations of oppositely directed exhaust jets have shown that MR signatures can extend over large distances \citep{gosling07b}. Presumably, the MR X-line can  be spatially extended to as long as hundreds of $R_E$ (Earth`s radius)  in the solar wind \citep{phan06}.

MR signatures have been observed mainly at the front or rear of MCs where the twisted field can in some cases be nearly opposite to the IMF. There is evidence that in such cases MR leads to flux rope erosion \citep{dasso06, lavraud14, ruffen12} which simply means that the outer layers of the MC flux rope are peeled off and the connectivity of the corresponding magnetic field lines is changed. Statistical analysis showed that  about 30\% of  MCs are associated with MR with equal distribution at the front and rear regions, leading to removal of 40\% of magnetic flux in average \citep{ruffenach2015}. This indicates that MR at the front and rear borders of MCs can significantly change the configuration of the magnetic field and therefore the geoeffectiveness of the ejecta. The boundary layers of MCs can be identified on the basis of the local occurrence of MR signatures, such as high proton temperature, proton density and plasma beta \citep{wei03}. However, MR does not always occur at the MCs boundaries along a spacecraft trajectory.
MR can also occur at thin current sheets in the solar wind which are not directly associated with MCs. In fact, using WIND spacecraft 3 s plasma and magnetic field data, narrow MR exhausts have already been observed in the turbulent high-speed solar wind \citep{gosling07a}.  Independently on the associated structures (ICME or turbulence generated), the local MR signatures in the solar wind include MR exhaust jets which are confined to a region of rotating magnetic field with decreased field amplitude, increased proton temperature and density, bounded by correlated (on one side of the exhaust) and anticorrelated (on the other side of the exhaust) velocity and magnetic field components. Usually a bifurcated current sheet is observed where the field reversal occurs in two large steps (two current sheets or Alfv\'en waves/discontinuities) at the leading and trailing edges of the exhaust. For the explanation of these signatures the Petschek model of MR is invoked which in fluid picture predicts an increased pressure as a result of separated plasmas entering to the exhaust from both sides. As the pressure increases, the pressure front propagates back against the incoming plasma flow, forming a pair of slow mode shocks, which locally decelerate and heat the plasma \citep{gosling12}. Although there exist slow shock observations associated with MR outflows in the solar wind \citep{sasunov12, zhou18}, the vast majority of exhaust boundaries are slow-mode-like only \citep{gosling12} where some of the signatures are absent, e.g. there is no electron heating associated with the slow shock. Moreover, the outlined exhaust boundary structure can be perturbed by generation of a turbulent boundary layer e.g. through Kelvin-Helmholtz instability \citep{sasunov12}, leading to plasma mixing, directional changes of magnetic field, vortices or current sheets \citep{voros14}.
The limited temporal resolution of the spacecraft and the huge scale ranges in the solar wind allowed us to understand MR physics mainly in the very context of magnetohydrodynamics (MHD).

Our understanding of the kinetic scale physics associated with MR comes predominantly from the  multi-point in-situ magnetospheric measurements, mainly from Geotail, CLUSTER, THEMIS and MMS missions. MMS, the Magnetospheric MultiScale  mission provided multi-point field, particle and plasma data with unprecedented high time resolution and revealed details of MR over ion and electron scales at the Earth's magnetopause \citep{burch16}, in the magnetotail \citep{torbert18, nakamura19}, within Kelvin-Helmholtz vortices at the magnetopause \citep{nakamura17} and in the turbulent magnetosheath \citep{yordanova16, voros17, phan18}.  Although MR is considered quasi-steady in the solar wind \citep{phan06}, the associated energy release in the magnetosphere or on the Sun is often explosive, reaching large-system-scales  and changing the connectivity of magnetic field lines from closed to open or vice versa \citep{zweibel09, treumann13, hesse20}. The large-scale consequences of MR observed in-situ at the Earth's magnetopause or in the magnetotail are the enhanced level of solar wind-magnetosphere coupling and substorms, respectively.  However, here the in-situ observations of MR events at different locations are complemented by multi-spacecraft observations in other locations or Earth-based (geomagnetic and radar observations). On the Sun, the large-scale visible consequences of MR, such as high temperature bright points 
\citep{peter14},
plasma inflow and outflow to the reconnection region 
\citep{yokoyama01}, acceleration of CMEs \citep{temmer08} are observed, rather than the small-scale violation of the ideal conditions.

It is expected that kinetic effects are important near the X-line \citep[eg.][]{burch16}, where both electrons and ions are demagnetized, or at separatrices \citep{lapenta15} and regions where the reconnection outflow is interacting with the ambient plasma \citep{lapenta17}.
At separatrices, parallel electric field accelerated particle beams, wave excitations \citep{khotyaintsev20}, instabilities \citep{hesse18}, density cavities \citep{lu10, norgren20}, double layers \citep{ergun09}, etc. can occur.

Since the majority of MR exhausts in the solar wind are crossed by a spacecraft far away from the X-line observations of kinetic signatures could be expected at reconnection outflow separatrices or at locations of outflow-ambient plasma interactions.
In fact,  kinetic effects associated with Hall magnetic field and two fluid reconnection physics have been observed 
far from the X-line in the solar wind \citep{huttunen08, xu15}. The boundary region of the outflows also corresponds to the location where enhanced high-frequency waves such as Langmuir waves, electron solitary waves and Doppler-shifted ion acoustic waves were observed by WIND/WAVES experiment \citep{huttunen07}. The observations of high-frequency waves indicate the presence of electron beams or current driven instabilities in outflow boundary regions. Since the burst mode plasma measurements from the WIND 3-DP experiment are available with 3 s  (spacecraft spin) temporal resolution only, high resolution wave observations are particularly important in evaluating the presence of electron scale interactions.

As can be seen from the outlined results of  in-situ, at times multi-point, and remote observations, collisionless MR in general is a fundamental multi-scale plasma physical process in which changes of magnetic field topology can lead to kinetic scale breaking of ideal frozen-in plasma conditions resulting in magnetic energy transformation to kinetic and particle energies. As a result, the newly reconnected field lines support plasma mixing and magnetic field topology changes, the associated electric fields accelerate particles, the reconnection outflow interacts with the ambient plasma generating turbulence and the free energy triggers instabilities and wave particle interactions. Although theory and numerical simulations support the outlined picture, none of these processes are fully understood. Specifically, in the solar wind, multi-point in-situ observations are mainly used in the context of space weather predictions rather than for observations of kinetic physics aspects or of large-scale magnetic field topology changes triggering MR, e.g. at the boundary layers of MCs. In general, neither the time resolution of plasma and field measurements are good enough to understand the relevant kinetic scale processes, nor the spacecraft crossings occur near the reconnection X-lines. Moreover, it is not known if the lessons on MR learned in the Earth's magnetosphere (mainly at the magnetopause and in the magnetotail) are applicable for solar wind MR which can be quasi-steady, fluid-scale (Petschek type) and 3D.

In order to understand better the multi-scale physics associated with MR at the boundary layer of a MC we re-examine the ICME event which occurred between 2000-10-02 and 2000-10-05, exploiting the available data from SOHO, ACE, GEOTAIL and WIND satellites.  MR occurring on 2010-10-03 within the boundary layer of MC has been studied in several papers \citep{gosling05, wang16, zhao19}. For this event, \citet{gosling05} found no substantial increase in the fluxes of MR associated energetic particles in 30-70 keV range. On the other hand, \citet{wang16} found that 100-500 keV electrons and 100 keV protons can be accelerated by MR  and the energetic particles were observed by WIND during the crossing of MR separatrix. Since the particles were observed at a large distance from the X-line, a clear physical explanation for energetic particle acceleration by MR was not offered. \citet{zhao19} explained the occurrence of MR  as a consequence of coalescence of two flux ropes, i.e. merging the MC  with a smaller size flux rope in the sheath. However, the alternative that the magnetic field topology in the sheath is the consequence of MR and plasma motions in front of the MC was not considered for this event.

The goal of this paper is to reveal the details of multi-scale processes occurring within the boundary layer of the MC by
\begin{itemize}
    \item closely looking at reorganizations of the magnetic field associated with MC and MR
    \item multi-point tracking of the motion of outflow channel and MR current sheet embedded into a large-scale flow shear
    \item examination of the outflow structure at both fluid and kinetic levels of description
    \item observation of particle acceleration and wave signatures
\end{itemize}
We argue that these observations are possible only because of the existence of the large-scale flow shear and the specific current sheet geometry in the studied event, which were not noticed in previous works.

The paper is organized as follows:
Section 2 describes the locations of the spacecraft and the available data products used for the event under study. The event overview is given in Section 3, where the large-scale structure of the ICME, the sheath and the MC extent and its interactions are explained. The MR outflow is located at the boundary of the MC and the freshly reconnected field lines are associated with plasma mixing. Section 4 is devoted to the more detailed description of ICME structure including the helical magnetic geometry of MC and the width of the sheath. It is shown here that MR outflow is located at the western flank of the MC. In Section 5 detailed observations of MR outflow and its interactions with the ambient plasma are described. MR outflow structure is studied in terms of both MHD description and kinetic physics. It is argued that the physical interpretation critically depends on the orientation of the MR current sheet.
Section 6 contains the summary and conclusions.

\section{Spacecraft locations and data}
The subplots a and b in Figure 1 show the locations of SOHO, ACE, Geotail and WIND spacecraft in the GSE coordinate system. GSE stands for the right handed
Geocentric Solar Ecliptic coordinate system in which the X axis points towards the Sun, the Z direction is perpendicular to the plane of the Earth's orbit, positive North.
Subplot c shows the magnitude of bulk velocity of the solar wind observed by the spacecraft where the jumps correspond to a fast forward shock driven by the MC under investigation. Based on both the shock normal directions at ACE and WIND and on the time delays of discontinuity passage between spacecraft, the shock front in X-Y plane is deviated from -Y axis  by 6 - 9 degrees in anti-clockwise direction.

The following data products are used:

From SOHO Mass Time-of-Flight (MTOF) proton instrument \citep{hovestadt95} solar wind proton bulk velocity is used with 30 s time resolution.

ACE magnetic field from Magnetic Field Experiment (MAG) \citep{smith98} is available with 1 s time resolution, plasma moments from Solar Wind Electron Proton Alpha Monitor (SWEPAM) \citep{mccomas98} with 64 s time resolution. Suprathermal electron pitch angle distributions are available from ACE SWEPAM-E.
The ionic charge composition of the solar wind is from the Solar Wind Ion Composition Spectrometer (SWICS) available with 1 h time resolution \citep{gloeckler98}.

Geotail magnetic field from Magnetic Field Experiment (MGF) \citep{kokubun94} is available with 0.063 s time resolution. Geotail Low Energy Particle (LEP) instrument, sensor unit Solar Wind (LEP-SW) \citep{mukai94}, observed ion moments with 1 min time resolution while the probe was at a larger distance from the bow shock.

The Magnetic Field Investigation (MFI) instrument on WIND \citep{lepping95} provides data with 0.092 s time resolution.
Ion  moments are used from the WIND 3-D Plasma and Energetic Particle Investigation \citep{lin95} with temporal resolution of 3 s. Ion and electron omnidirectional fluxes electron pitch angle distributions are available with time resolution of $~$24 s. Electron moments from the Solar Wind Experiment (SWE) instrument \citep{ogilvie95} are used with 12 s time resolution (onboard calculated) and with $\sim$98 s time resolution (ground calculated). Electron density, plasma wave and electric field data in kHz range are used from the WAVES investigation on the WIND spacecraft \citep{bougeret95}.

\section{Event overview}
Figure 2 is the overview plot of the large-scale structure of ICME, observed by ACE spacecraft between 2000-10-02 22:00 and 2000-10-05 06:00 UT. The top two subplots a and b are feather plots showing the magnetic field vectors with GSE components $B_{XY}$ and $B_{XZ}$ in equally spaced points along the horizontal time axis. The time step between the consecutive points is 1000 s, which corresponds to $\sim$ 17 min. The feather plots show direction and magnitude of the magnetic components along the time axis. The meaning of the feather plots becomes clear when they are compared to the GSE magnetic field components in subplot c. Subplots 2 d, e show the GSE $V_X$ and $V_y$, $V_Z$ proton velocity components, respectively. Proton density $N_p$ and temperature $T_p$ are shown in the subplot 2 f. The supra-thermal electron pitch angle (e-PA) distribution is depicted in the subplot 2 g. The bottom subplot 2 h shows the oxygen $O^{7+}/O^{6+}$ ratio with 1 h time resolution.

It is easy to recognize the well-known above described general signatures of MCs in Figure 2. During the MC time interval labeled  by the horizontal light-brown thick lines, the feather plots (subplots 2a,b) and the magnetic field components (subplots 2c) exhibit smooth helical rotations, $T_p$ is low and fluctuates around 1 eV (subplot 2f), bi-directional suprathermal electrons occur (subplot 2g: parallel and anti-parallel electrons relative to the magnetic field) associated with increased $O^{7+}/O^{6+}$ ratio (subplot 2h).

It can be seen from the magnetic data, from the feather plots (Figures 2a, b) and magnetic components (Figure 2c), that during the first half of the MC the magnetic field rotations are mainly in $B_Y$ component which changes sign. Additionally $B_Z$ first increases then decreases, however without a sign change. During the second half the rotations are  in $B_{XZ}$ components, where $B_Z$ changes sign. Obviously, GSE  is not the physically relevant coordinate system of the MC.

MR occurs at the boundary of MC indicated by red dashed lines in Figure 2. The reconnection current sheet is characterized by sudden directional changes of the magnetic field (subplots 2a-c) associated with enhanced velocity components (outflow, subplots 2d, e) and enhanced density and temperature. The MC drives the fast forward shock S1 indicated by the dashed vertical line shortly after midnight on October 2.  Between the shock S1 and MR/MC boundary the ICME sheath region (indicated by black horizontal thick line in subplot 2a) with enhanced magnetic fluctuations (subplot 2c) exists. The fast forward shocks S2 and S3 (vertical dashed black lines) occur at the rear of the MC, both associated with jumps in magnetic field magnitude, solar wind velocity, density and temperature. The compression of MC by other ejecta or the propagation of shocks through a MC can result in more complex structures generating extreme values of the southward component of the magnetic field ($B_Z<$0) and a potential geomagnetic impact \citep{lugaz17,werner19,dimmock19}. In our case, in fact, there is a long duration of negative $B_Z$ associated with S2 and S3 (subplot 2c).
As  seen from subplot 2g, all three shocks are associated with pitch angle scattering. At shocks S1 and S2 the largest electron fluxes are in field-aligned direction.

Although there are no direct signatures for MR (e.g outflows or current sheets) in the rear of MC, the
$O^{7+}/O^{6+}$ ratio remains high there, decreasing towards the shock S3 (subplot 2h), indicating possible mixing along reconnected field lines. Another indirect signature for MR is the occurrence of gaps in the anti-parallel electron pitch angle distribution inside the MC between 01:00 and 11:00 UT on October 04 (subplot 2g). The missing bidirectional electron distributions indicating the presence of open field lines inside the MC do not necessarily mean that the MR X-line is near the crossing trajectory of spacecraft.

In this paper we  thoroughly investigate  the multi-scale physical processes associated with MR at the front boundary of MC (red dashed vertical lines). This region is far from the shocks, located at the sharp boundary of the first appearance of bidirectional electrons, which is actually the boundary layer of the MC (subplot 2g).

There is also a layer attached to MR in the sheath marked by blue vertical dashed lines, we call it mixing layer (ML). This layer might result from at least two different physical processes. One is MR at the MC boundary generating freshly reconnected MC - IMF field lines leading to mixing of plasma in the reconnected flux tubes. As a consequence, the cold plasma in the MC is mixing with the hotter plasma in the sheath with a transition from $T_p\sim$1 to 9 eV in ML (subplot 2f). At the same time, there is a leakage of $O^{7+}/O^{6+}$ rich plasma from MC to ML. This ratio is reaching 0.6 in ML, while it fluctuates around 1 in MC and is around  0.2 in the sheath (subplot 2h).
The other process occurring near the boundary of MC is the specific way of interaction of sheath plasma with the reconnected field lines. The details of this interaction will be explained later, here we  point out the field and velocity changes happening in the ML. From the front to the rear of MC $|V_{X}|$ is decreasing from 430 to 390 km/s. In the ML the high speed of the flow in the sheath downstream of the shock decreases from $|V_{X}| \sim $450 to $\sim$ 410 km/s (subplot 2d). Moreover, the $B_{X}$, $B_{Y}$ field components remain unchanged, $B_{Z}$ increases from -1 to 8 nT (subplot 2c), $V_{Z}$ decreases from 60 to 15 km/s and $V_{Y}$ increases from 0 to 15 km/s. Another striking feature is the existence of a large-scale flow shear seen in $V_Z$ across ML and MR between 12:00 and 22:00 UT on October 03. We note that, between 2000-10-02 22:00 and 2000-10-05 06:00 UT, WIND observes  similar field, plasma and electron pitch angle signatures as ACE (not shown).

\section{ICME structure}
In this Section we investigate the MC driven shock S1, the ICME sheath width, the helicity, axis orientation, the observed duration (beginning and end) of the MC flux rope along the crossing trajectories of the spacecraft. Instead of fitting a flux rope model, the emphasis is on understanding of the MC geometry needed for the interpretation of MR observations.

\subsection{MC flux rope helicity and axis}

The sharp boundary of the first appearance of bidirectional electrons and closed magnetic field lines immediately after MR outflow  can be considered to be the MC boundary (Figure 2). Although an MC boundary is  a boundary layer of finite width, rather than a sharp border, the localized outflow structure seen in time by each spacecraft can serve as a reference point for further analysis.

Figure 3 shows the magnetic field observations by ACE, Geotail and WIND between 13:00 UT October 03 and 22:00 UT October 04. Geotail and WIND data have been time-shifted to match ACE observation of MR (MC boundary).
In this way the geometry of MC observed by different spacecraft can be compared. Subplots 3a-c show magnetic observations in the GSE system. Minimum variance analysis (MVA) for each spacecraft was used to transform the GSE magnetic field components to the MC  coordinate system. Initially the time interval marked as MC in Figure 2 was used for MVA. Start and end times were slightly changed to find the optimal time interval giving the best MVA eigenvalue ratios. Subplots 3d-f show the transformed magnetic field to maximum ($Bmax$), intermediate ($Binterm$) and minimum ($Bmin$) variance components for each spacecraft. The sign-changing component is $Bmax$. Comparing the GSE and MVA magnetic field components it becomes clear that $Bmax \sim B_Y$ and $Binterm \sim B_Z$, however, $Binterm$ remains positive in MC. Neglecting deformations of MC and supposing a cylindrical shape, subplot 3g explains the geometry of the flux rope in the $Bmax - Binterm$ plane during the crossing of the structure by spacecraft (red dashed line). The minimum variance direction corresponds to the direction of MC axis and $Bmin \sim B_X > 0$ nT, i.e. the axis points roughly towards the Sun. From subplot 3g is clear that the MC has a left-handed helicity. The MC axis direction determined along WIND trajectory in GSE coordinates is [0.94 -0.11 -0.32], inclined at an angle of 19 degrees from the GSE X axis in X-Z plane (subplot 3h) and at a declination angle of 7 degrees from the GSE X axis in X-Y plane (not shown). MC axis orientation from Geotail is by 12 degrees and from ACE is by 13 degrees different from WIND MC axis, leading to inclination angles 30 and 32 degrees from GSE X direction, respectively. The end of the MC or helical flux rope can be seen after 10:30 UT on October 04, when two components of the MVA field,  $Binterm$  and $Bmin$ starts fluctuating around 0 nT.
ACE density data (subplot 2f) show that at the end of the MC and before the shock S2 the density is decreasing which together with the magnetic field signatures $Binterm, Bmin \sim$ 0 nT (subplots 3d-f) indicate that the spacecraft is crossing the wake behind the MC. The less dense solar wind  is known to change the propagation conditions for other structures in the wake of ICMEs \citep[e.g.][]{rollett14}. Since the shock S2 is faster than the MC (subplot 2d) the wake at ACE is wider than at Geotail or WIND (subplots 3d-f). Subplot 3h shows the crossing of the spacecraft across the helical MC, $\sim$5000 $R_E$ wide, to the wake region. Due to the inclination of the MC axis and the -$V_Z$ velocity (Figure 2e) the spacecraft do not observe more helical structures of the MC.

\subsection{Width of the ICME sheath}
Using the MR outflow right boundary (at the MC boundary) as a reference point the distance of MC boundary to the MC driven shock S1, which is the width of the sheath along a spacecraft trajectory, can also be determined from the data. Figure 4 shows the magnetic field (magnitude and $B_X$), $|V|$ and density (N) observations by ACE (subplots 4a-c), Geotail (subplots 4d-f) and WIND (subplots 4g-i). Again, Geotail and WIND data have been time-shifted to match ACE observation of MR (blue vertical dashed line, MC boundary). Because of the time-shift the S1 shock is not observed at the same time as it is in Figure 1c. Although the plasma data from Geotail are missing at MR location, the boundary can be identified from the magnetic data (subplot 4d). The shock S1 seen by the spacecraft can be readily identified from magnetic and plasma data (vertical green and red dashed lines near midnight). Since ACE and Geotail are relatively close in GSE Y and Z coordinates (Figure 1), the shock is observed by these spacecraft almost at the same time (green and red vertical dashed lines). The time difference between the shock observations is $~$9 minutes.  The width of the ICME sheath is roughly the same along the trajectories of ACE and Geotail. However, WIND is located more than 220 $R_E$ in -Y GSE direction from ACE location and it observes the (time-shifted) shock about $\sim$50 minutes earlier (brown dashed vertical line) than Geotail suggesting that the ICME sheath is wider along the WIND trajectory. This might not be true if the local velocities of MC boundary or/and shock were too different. As for the shock the ACE and WIND shock database
compiled by J. Kasper (Harvard‐Smithsonian Center
for Astrophysics; \break
http://www.cfa.harvard.edu/shocks/) was used, where several shock analysis
methods were applied to determine the shock parameters. Median S1 shock normals in GSE are $\mathbf{n}_{ACE}$=[-0.97 -0.16 -0.2] and $\mathbf{n}_{WIND}$=[-0.89 -0.09 -0.4]. Due to errors
$n_{err}$=[$\pm$0.04 $ \pm$0.1 $ \pm$0.3] the normals are indistinguishable. Also, the shock velocities are very similar, $V_{sh-ACE} = 464 \pm 4$ km/s and $V_{sh-WIND} = 462 \pm 9$ km/s. The estimated velocity of the MC boundary based on 0.5 hour average after MR  (vertical blue line) is $V_{MC} = 430$ km/s (subplots 4b, h). The radial MC expansion speeds at ACE and WIND, calculated through $V_{exp}=(V_{LE}-V_{TE})/2$ (the subscripts LE and TE refer to leading and trailing edge MC speeds, respectively) are also comparable. The mean expansion speed at ACE is $V_{exp}=$26, at WIND is $V_{exp}=$22 km/s, while the deviations from the mean are $\pm$8 km/s.
Since the shock speeds and the MC expansion speeds are practically the same at ACE/WIND positions, we interpret the time difference between the observations of the forward shock by these spacecraft, relative to the MC boundary (Figure 4), as a broadening of the sheath along ACE and WIND crossings. Converting the time difference by using the average solar wind speed, the ICME sheath is found to be wider along  WIND trajectory by $\sim$220 $R_E$ than along the trajectories of Geotail or ACE.

\subsection{ICME large-scale geometry}
Summarizing the previous findings, such as the deviation of the S1 shock front from GSE Y direction (Figure 1a), the MC axis direction approximately towards the Sun, and the width of the ICME sheath along spacecraft paths, we find that the crossings occur at the western flank of the ICME.  Without going to further details described in subsequent steps below we refer to Figure 10a which shows the large-scale geometry of the ICME, shock and left-handed MC, together with the trajectories of the spacecraft in GSE X-Y plane. MR in the interplanetary space is usually considered to occur in front or rear of MCs when the the flux rope axis lies largely in the transversal GSE Y-Z plane \citep{janvier13, ruffenach2015}. However, magnetohydrodynamic simulations of MCs has shown that, depending on the magnetic helicity of the flux rope and on the orientation of IMF, MR can occur at the flanks of MCs as well \citep{taubenschuss10}.

\section{Observations of MR}
In this section MR is studied in both GSE and reconnection LMN coordinate systems. The goal is to understand the propagation and evolution of MR outflow channel between the spacecraft. It will be shown that for understanding of the MR outflow structure and  interactions of the outflow with the ambient plasma both fluid and kinetic scale descriptions are needed.

\subsection{MR in minimum variance (MVA) coordinates}

It is non-trivial to determine the coordinate system in which MR outflow structure can be properly described. First we use the MVA coordinate transformation which transforms the variables from GSE to LMN coordinates, where $\mathbf{L}$ is the direction along the outflow, $\mathbf{N}$ is the normal direction to the current sheet, $\mathbf{M}$ is the "out-of-plane" direction along MR X-line. Then, on the basis of physical considerations, the $\mathbf{M}$ and $\mathbf{N}$ directions will be rotated around $\mathbf{L}$. We will call this coordinate system as "MVA-rotated".

The reconnection current sheet was observed by ACE, Geotail and WIND spacecraft. Plasma data are not available from Geotail during current sheet crossing. In SOHO solar wind bulk velocity data the MR outflow was not observed (not shown).
Figure 5a shows the GSE magnetic field components from ACE.  Figure 5b shows the transformed $B_L$, $B_M$ and $B_N$ components of the magnetic field, with a typical two-step sudden changes of $B_L$ at red and black dashed vertical lines, indicating that the current sheet is bifurcated. This event has already been studied by \citet{wang16} and \citet{zhao19}. In both papers the same $B_{LMN}$ magnetic field components were presented as in Figure 5b.  In this coordinate system $B_M \sim$ 7.5 nT before and after the current sheet, indicated by horizontal dashed black line in Figure 5b. This would suggest that there is a guide field ($Bg$) in out-of-plane direction, i.e. the reconnecting magnetic field lines are not fully anti-parallel in the L-M plane. Across the current sheet (bordered by vertical dashed lines)  B$_M$ $>$ Bg and the B$_M$-Bg difference gives the Hall magnetic field $B_{Hall}$. The Hall magnetic field appears as a result of differential motion between electrons and ions (current) preferentially near the X-line within the ion/electron diffusion regions. In the solar wind the Hall magnetic field has been observed far, possibly thousands of ion inertial lengths (di) away from the X-line  \citep{xu15, mistry16}, associated with $\mathbf{L}$  directional currents near separatrix. In large distances from the X-line $B_{Hall}$ could be generated by currents flowing in separatrices. However, B$_M$-Bg is a smooth curve in Figure 5b. Although a strong Bg could generate a roughly unipolar $B_{Hall}$ far from the X-line, however, many MR events with significant guide fields do not have Hall fields in the solar wind \citep{mistry16}. Therefore,
the MVA coordinate system should undergo further tests. Since MVA is performed across the whole current sheet, including the bifurcated currents with significant rotations of the magnetic field, there is no guarantee that minimum variance direction corresponds to the $\mathbf{N}$ direction, especially in a 3D system. The eigenvalue ratios $\lambda _1/\lambda _2$ and $\lambda _2/\lambda _3$ of the MVA variance matrix corresponding to maximum ($\lambda _1$), intermediate ($\lambda _2$) and minimum ($\lambda _3$) variance directions have been calculated for different nested time intervals across the current sheet. For time intervals including the outflow and additional 10 minutes before and after the outflow, boths eigenvalue ratios are larger than 10. As the length of the time interval comprising the whole current sheet shortens $\lambda _1/\lambda _2$ remains over 5, however, $\lambda _2/\lambda _3$ drops below 2. This indicates that the MVA variance ellipsoid can be cigar shaped with axis in $\mathbf{L}$ direction. This is called MVA degeneracy, when the
$\mathbf{M}$ and $\mathbf{N}$ directions are in a plane perpendicular to $\mathbf{L}$, however, pointing to arbitrary directions \citep{sonnerup98}. In what follows we will refer to the LMN coordinate system obtained for the wider  time interval and used also in other papers for this event \citep{wang16, zhao19} as the MVA system. The tag "MVA rotated" will be used for the coordinate system obtained from MVA by rotation of  the $\mathbf{M N}$  eigenvectors around $\mathbf{L}$  by 90$^{\circ}$ in clockwise direction  or by other angles.

It is easy to notice that the $Bg$ guide field can be transformed out by rotating of the MVA coordinate system. The $\mathbf{M N}$ eigenvectors were gradually rotated around  $\mathbf{L}$ in order to find a configuration which minimizes $Bg$. However, the effect of different rotations on the physical parameters was also tested.   The magnetic field components obtained by counter-clockwise rotation of $\mathbf{M}$ and $\mathbf{N}$ eigenvectors around the $\mathbf{L}$ direction for the three spacecraft are shown in Figures 5c-e. Interestingly, when after the coordinate rotation $Bg$ becomes $\sim$ 0 nT  at ACE and Geotail, also the large $B_{Hall}$ field disappears and smaller fluctuations remain only (Figures 5c,d). There is a small negative Bg at WIND and more pronounced B$_M$ fluctuations near the current sheet boundaries (Figure 5e). Otherwise, ACE, Geotail and WIND observe very similar current sheet profiles. The bifurcated current sheet is readily discernible in all three cases. The right boundaries of the current sheets, between the dashed black and blue vertical lines  are a bit different.

Since the guide field and the Hall field of kinetic origin can significantly change the outflow structure and dynamics it is important to determine which coordinate system is the valid one.
First, the $\mathbf{L}$ direction is tested.
Figure 6a shows the locations of the spacecraft together with the MVA direction $\mathbf{L}$ in the GSE Y-X plane. Here the differences of the locations of the spacecraft in GSE Z direction are neglected. The $\mathbf{L'}$  direction corresponds to the direction of perpendicular to magnetic field $\mathbf{L}$  directional outflow (see later). The average angle between $\mathbf{L}$  and $\mathbf{L'}$  directions between spacecraft is less than 10 degrees. Both $\mathbf{L}$  and $\mathbf{L'}$  directions at ACE are projected to the upstream GSE Y location of the WIND spacecraft. Supposing plane geometry for the MR outflow channel, average solar wind speed of $V_{SW} = 430$ km/s, the arrival time from upstream points W' and W (X GSE = 490 and 400 $R_E$) to the WIND location (X GSE = 32.8 $R_E$) can be predicted (Figure 6a). The predicted times are 113 and 91 minutes, respectively, while the measured arrival time of the outflow structure  from ACE to WIND is 108.6 minutes. On this basis we find that the MVA $\mathbf{L}$  direction is satisfactory. We further investigate the already above described rotation of the coordinate system around $\mathbf{L}$  direction. Notice, that the averaged distance between ACE location and W', W points along the outflow is 314 $R_E$. Since the temporal profiles of the current sheet at ACE, WIND and Geotail are almost the same (Figure 5), the distance to the MR X-line is $\gg$ 314 $R_E$.

Figures 6b and 6 c.1 show the difference between MVA and MVA-rotated LMN coordinates relative to XYZ-GSE  and magnetic field directions.
The MVA and MVA-rotated eigenvectors for each spacecraft are in Table 1.

In the MVA system (Figure 6b right), the reconnection out-of-plane $\mathbf{M}$  component points roughly into Z GSE direction, the $\mathbf{L, N}$ components lay roughly in the X, Y  GSE plane. This means that the current sheet LN plane is crossed by the spacecraft along the direction of the solar wind bulk velocity, which is $V_X\sim$ -430 km/s (Figure 6b left). The calculated width of the outflow channel along $\mathbf{N}$ is 244000 km ($\sim$12 minute crossing time along X GSE, Figure 5) which is 3050$d_i$, where $d_i$=80 km is the average proton inertial length during the outflow. The wedge angle of MR outflow in the LN plane is usually determined as an angle between the normals of the discontinuities at the leading and trailing edges of the current sheet. Due to boundary fluctuations, however, the normals and the distance to the X-line cannot be determined.
Comparing Figure 6b with Figure 5b it can be seen that, before crossing the current sheet $B_L \sim$ 15 nT, after the current sheet $B_L =$ -14.5 nT and there is small negative average $B_N$ within the outflow. The crossing is approximately perpendicular to out-of-plane direction $\mathbf{M}$, along X GSE. Considering the high crossing velocity, the spacecraft are a very short time near MR separatrices only. Now we define $\mathbf{B1}$ as an average magnetic field vector before current sheet crossing determined in a 10 minute long time interval outside of the immediate vicinity of the current sheet where boundary fluctuations are enhanced. The average $\mathbf{B2}$ is defined in a similar way after the current sheet.
The vectors $\mathbf{B1}$  in Figures 6b,c are always the first observed magnetic field vectors in time. Near the ecliptic plane the spatial location of  $\mathbf{B1}$ is roughly in negative -X  GSE direction from the location of  $\mathbf{B2}$ (Figure 6b). In vertical case (Figure 6c.1) the spatial location of $\mathbf{B1}$ is roughly in positive Z GSE direction from the location of  $\mathbf{B2}$.

For now we keep the MVA-rotated coordinate system as it is described above ( Figure 6c.1), however, we will further consider the possibility that  $\mathbf{M}$  and $\mathbf{N}$   are rotated around the $\mathbf{L}$  direction in clockwise or counter-clockwise directions using a rotation matrix $\mathbf{ R}$ (case c.2 in Figure 6c).
When the rotation is by 90$^{\circ}$ in counter-clockwise direction,  the MVA rotated coordinate system is identical with the MVA system.

\subsubsection{The vertical motion of the MR outflow channel between the spacecraft}
Supposing a vertically oriented current sheet (Figure 6c.1), here we provide an order of magnitude predictions of the motion  of MR outflow channel between the spacecraft.

In the MVA-rotated system, the reconnection LN plane is approximately perpendicular to the ecliptic, therefore, the crossing of the current sheet requires a vertical $V_Z$ speed (Figure 6c left). The positive $V_Z$ of the ambient plasma surrounding the MR outflow can already be seen in Figure 2e.
At the same time the current sheet is convected by the solar wind bulk speed ($\sim V_X$). Along the outflow in the spacecraft frame, the outflow front speed is the sum of the solar wind bulk speed and the outflow speed. Here we should be able to answer the question why the vertical current sheet with $\mathbf{N}$  $\sim$ Z GSE undergoing a vertical motion in positive Z GSE direction can be seen by ACE, Geotail and WIND, but not by SOHO. To show that we will use the time-delay of outflow observations between the spacecraft, calculate the spatial vertical shift of the supposedly planar outflow channel with constant average velocity $V_Z$
and estimate an additional change in  the vertical location of the flow channel in a given position which appears due to the inclination of $\mathbf{L}$  direction relative to the GSE-XY plane. The latter can be positive or negative depending on the position of a given location relative to a reference spacecraft. Although the $\mathbf{L}$  direction and the inclination angle slightly change between the spacecraft, initially we choose constant values for these quantities  between the spacecraft.

Figure 7 shows the time delays in MR outflow observations (all in GSE) between ACE (Figures 7a-e), Geotail (Figure 7f) and WIND (Figures 7g-k) spacecraft.  For each spacecraft the magnetic field magnitudes are shown in Figures 7a, f and g. The $B_X$, $B_Y$ and $B_Z$ components are in Figures 7b and h. The magnitude of  radial $|V_X|$ and transverse $V_{Y,Z}=\sqrt(V_Y^2+V_Z^2)$ velocities are in Figures 7c and i. The $V_Y, V_Z$ solar wind velocity components and ion density Ni are in Figures 7d and j. The ion density and temperature are in Figures 7e and k. In this panel the same Y axis is used for Ni and Ti. For this time interval there are no reliable plasma measurements from Geotail in the solar wind.

The sharp boundary in $|B|$ at the right side of the outflow at each spacecraft can be used as a reference point to estimate the time delay between outflow observations. At ACE the sharp boundary in $|B|$ is seen at 15:09 UT (Figure 7a). The same structure is seen by Geotail  t(ACE-Geotail)=50.2 minutes later (Figure 7f) and by WIND t(ACE-WIND)=108.6 minutes later (Figure 7g). The large-scale shear in $V_Z$, which was explained in Figure 2e, is seen here  at both ACE and WIND (Figures 7d, j). However, at ACE, the outflow is embedded into a flow structure with $V_Z \sim$ 18 km/s (Figure 7d). Since the ambient magnetic field  before and after the outflow is predominantly in the X-Y GSE plane (Figure 7b), the $V_Z$-GSE velocity is approximately perpendicular to the magnetic field there.
The vertical crossing speed and the duration of the outflow gives for the current sheet thickness 12900 km ($\sim$ 2 $R_E$ or 157$d_i$).

The vertical motion of the outflow channel from ACE GSE Z = -7 $R_E$  (Figure 1) to Geotail location can be predicted.
Supposing planar geometry for the outflow and a constant $V_Z$ between the spacecraft, the vertical shift of the flow channel is 8.5 $R_E$, i.e. the outflow structure moves from ACE GSE Z = -7 $R_E$  to the location Z GSE $\sim$ 1.5 $R_E$ (not shown). This value has to be corrected by $\Delta$Z which is due to the inclination of the $\mathbf{L}$  direction relative to the GSE-XY plane. Since at Geotail $\mathbf{L}$ =[-0.63 0.78 -0.06], the flow inclination angle is 3.4$^{\circ}$ and over the distance of $\Delta L$ along $\mathbf{L}$  (Figure 6a), $\Delta Z$ =1.3 $R_E$. By adding it to 1.5 $R_E$, the predicted vertical location of the flow channel is Z GSE $\sim$2.8 $R_E$ at Geotail. This roughly matches the actual Geotail Z GSE = 2 $R_E$ location (Figure 6a), where the outflow is de-facto observed. The 0.8 $R_E$ difference can appear as a result of changing $V_Z$ between the spacecraft by 1-2 km/s or/and due to the error in determination of L inclination angle by $<$1$^\circ$.

SOHO at GSE [X,Y] = [241,89] $R_E$ should see the $\mathbf{L}$  directional outflow at the same time as it goes through an upstream location A1=GSE [X,Y] = [226+$\Delta$A1,-30] $R_E$, where [226,-30] $R_E$ is the location of ACE (Figure 1) and $\Delta$A1=110 $R_E$ is determined by triangulation (not shown).  On this basis SOHO should see the outflow $\sim$1630 s earlier than ACE. During this time the vertical shift of the outflow structure due to $V_Z$ = 18 km/s is $\sim$4.6RE.  At ACE $\mathbf{L}$ =[-0.62 0.78 -0.08] and the inclination angle of $\mathbf{L}$  is 4.6$^\circ$.  This gives over a distance of $\Delta$L=152 $R_E$ (distance between upstream location A1 and SOHO, determined by triangulation) a positive shift $\Delta$Z=12 $R_E$, which adds to 4.6 $R_E$, leading to the predicted vertical location of the outflow of Z GSE =16.6 $R_E$, while the actual SOHO Z GSE =-13 $R_E$ (Figure 1).  Since the thickness of the current sheet is $\sim$ 2 $R_E$, this might explain why SOHO does not observe the outflow (not shown).

The prediction of the vertical shift of the outflow channel from ACE to WIND is connected with more uncertainties because of the large GSE-XY differences in their locations.
ACE at GSE [X,Y] = [226,-30] $R_E$ should observe the $\mathbf{L}$  directional outflow at the same time as it goes through an upstream location $W$=GSE [X,Y] = [32.8+$\Delta W$,-252] $R_E$, where [32.8,-252] $R_E$ is the location of WIND (Figure 6a) and $\Delta W$= 367.2 $R_E$ is determined by triangulation. We consider here only the upstream point W at GSE [X,Y] = [400,-252] $R_E$ (Figure 6a). Since WIND observes the outflow 108.6 minutes later than ACE (Figure 7) the vertical shift of the outflow channel due to $V_Z =$ 18 km/s during this time is $\sim$ 18.4 $R_E$. At ACE, Geotail and WIND the $\mathbf{L}$  directional inclination angle relative to the GSE XY plane varies as 3.4$^\circ$, 4$^\circ$ and 4.6 $^\circ$, respectively. By taking the smallest inclination angle, this gives over a distance of $\Delta L$=279.5 $R_E$ (here this is the distance between upstream location $W$ and ACE, determined by triangulation)  $\Delta Z$=-16.5 $R_E$, which adds to +18.4 $R_E$ (due to $V_Z$=18 km/s), leading to a net vertical shift of the flow channel by +1.9$R_E$. Thus, the flow channel from ACE GSE Z=-7 $R_E$ is predicted to shift to GSE Z=-5.1 $R_E$ at WIND while the actual WIND is at
GSE Z=-4 $R_E$. By taking $V_Z$=19 km/s between upstream point W and WIND, the net vertical shift is modified to +2.9 $R_E$, leading to the predicted vertical location of the flow channel of -4.1 $R_E$ at WIND which is matching WIND  GSE Z=-4 $R_E$ location pretty well.
The above rough estimations of the vertical motion of MR outflow channel (the orientation of the LN plane is shown in Figure 6c.1) support the idea that the relatively thin flow structure (of width $\sim$ 2 $R_E$) can be sequentially observed by ACE, Geotail and WIND spacecraft, but not seen at SOHO location.

\subsection{MR outflow structure}
The analysis of the MR outflow structure is divided into several steps. Since the outflow can interact with the ambient plasma and magnetic field, the vicinity of the exhaust containing flow shears and discontinuities will be considered first. This has already been partially touched upon in previous sections mentioning the large-scale mixing and the motion of freshly reconnected field lines in the ICME sheath. Then the fluid description of the outflow channel, also with regard to the interactions with the ambient plasma will be presented. Finally, kinetic effects capable of shaping the outflow structure will be presented.

\subsubsection{The vicinity of MR outflow - slow-shock-like discontinuities}
First the density structure around MR outflow is considered. The yellow dashed lines in Figures 7a-e (ACE) and in Figures g-k (WIND) show that at the density jumps, $|B|$, Ti and velocity components exhibit smaller or larger jumps. From Geotail $|B|$ is shown only (Figure 7f).

The jumps in different parameters supposedly correspond to discontinuities.  Before the outflow in the ICME sheath (at vector $\mathbf{B_1}$ in Figure 6) the discontinuities are labeled as D1A (roughly at 14:12 UT), D1G (at 15:00 UT) and D1W (at 16:35 UT), as observed by ACE, Geotail and WIND, respectively. The discontinuities after the outflow at the boundary layer of the MC (at vector $\mathbf{B_2}$ in Figure 6) are marked similarly as D2A (at 15:38 UT), D2G (at 16:30 UT) and D2W (at 17:18 UT). Although there might be other discontinuities earlier or later in time, we are interested in those when $|B|$ decreases and Ni changes before and after the outflow. Here our aim is not to verify the Rankine-Hugoniot jump relations \citep[e.g.][]{burlaga95}, but  rather to understand possible interactions between discontinuities and the outflow. The other goal is to compare the evolution of discontinuities on the basis of available data at different distances from the MR X-line. The field and plasma parameters across discontinuities D1A and D1W in the ICME sheath are in Table 2. The same for D2A and D2W is shown in Table 3. The GSE field and plasma variables represent 10 minute averages before and after the discontinuity, not including the largest gradients at jumps. The downstream side of a discontinuity is where $|B|$ decreases, Ni and Ti increase. The discontinuity MVA normals in GSE are shown in the first rows in Tables 2 and 3.

Across D1A (Table2, left) $|B|$ decreases, $V$ decreases, both Ni and Ti increase. $V_X$ does not change, the transverse ion velocities $V_Y$, $V_Z$ and $V_{YZ}=\sqrt{V_Y^2+V_Z^2}$ (Figures 7c and d) decrease. Since the local Alfv\'en speed is $\sim$ 100 km/s the velocity changes are sub-Alfv\'enic.
Initially we suppose that, D1A is neither a forward slow shock (SS), for which Ni, Ti and V should increase, B decrease nor a reverse SS, for which V and B should increase, Ni, Ti decrease. Across D1A the total pressure (magnetic and ion thermal pressure, electron thermal pressure is not available) changes from 0.15 to 0.1 NPa (not shown). However, without the electron thermal pressure it is not possible to classify D1A as a pressure balanced structure. Since $|B|$ and Ni change across D1A, it is not a rotational discontinuity. Although in anisotropic plasmas $|B|$  and Ni could change across a rotational discontinuity, for low beta plasmas (in our case beta $<$ 0.1) Ni could change by at most a factor of 1.1 and $|B|$ should be almost constant at 1 AU \citep{hudson73}. In our case Ni changes more than 1.6 times.  D1A could also be  a tangential discontinuity. In that case, supposing pressure balance, Ni, Ti and $\mathbf{ B}$ can change across D1A, however, the normal component of $\mathbf{ B}$ should vanish at the discontinuity surface.
By calculating the scalar product $\mathbf{B}_{u,d}\cdot \mathbf{n} \sim$ 6 nT,
we find that the normal component does not vanish at D1A. The subscripts u and d refer to the upstream and downstream values, $\mathbf{n}$ is the normal.

D1A can be considered also as a velocity stream interaction region separating faster, less dense and cooler plasma from slower, denser and a bit hotter plasma. Obviously, this is not a solar wind stream interaction region separating a fast stream of coronal hole origin from a slow solar wind \citep[][e.g.]{richardson18}, but rather, a velocity shear generated by the moving and possibly expanding MC interacting with the ambient plasma in the sheath. Part of this interaction happens at the MC boundary where the MR outflow is moving faster than the MC. The denser and hotter plasma carried by reconnected open field lines piles up and slows down at D1A stream interface (Figures 7a-e, Table 2 left). Actually, this could be an SS if the velocity was not increasing towards the upstream direction. However, the vertical flow shear seen in $V_Z$ and also to a less extent in $V_Y$ (Figure 7d and also in Figure 2e) can mask the real velocity jump conditions across D1A. This 'masking' does not allow to calculate reliably the SS speed. For this reason we identify D1A as an SS-like discontinuity embedded into the flow stream. The angle $\Theta_{Bn}$ between $\mathbf{B}$ and  $\mathbf{n}$ is $\sim$67 $^\circ$ which corresponds to a quasi-perpendicular shock geometry.

Geotail basically observes the same $|B|$ profile at D1G as ACE at D1A (Figure 7f). The time delay between the discontinuities D1A, D1G and the left border of the current sheet is $\sim$ 48 minutes (Figures 7a, f). Since Geotail is almost at the same GSE Y as ACE (Figure 6a), the distance to the X-line at both spacecraft presumably is not very different, which can result in similar $|B|$ profiles. D2A is also roughly at the same distance from the right boundary of the current sheet as D2G, though D2G is less pronounced. Since plasma data are not available from Geotail, the D1G and D2G discontinuities are not analyzed further.

Wind observes D1W only 11.5 minutes earlier than the left border of the outflow (Figures 7 g-k).
Across D1W (Table 1, right) B decreases, $V$, Ni and Ti increase. As for velocities mainly $|V_X|$ and $V_Z$ increase. The normal components of the magnetic field are $\sim$ 15 nT, and  $\Theta_{Bn} $= 12$^\circ$ and 4$^\circ$, respectively. These are the signatures of a quasi-parallel forward SS.
The slow shock speed normal to the shock surface is estimated from the upstream and downstream values as $V_{sh} = \mathbf{n} [N_d\mathbf{V}_d \cdot \mathbf{n} - N_u\mathbf{V}_u \cdot \mathbf{n}]/(N_d-N_u) $. Then the normal solar wind speeds in the shock frame are calculated as ${U_n}_u= \mathbf{V_u} \cdot \mathbf{n} - V_{sh}$ and ${U_n}_d= \mathbf{V_d} \cdot \mathbf{n} - V_{sh}$, respectively. Finally, the Alfv\'en-Mach number is estimated through ${M_A}_u = {U_n}_u/(V_A cos\Theta_{Bnu})$ and ${M_A}_d = {U_n}_d/(V_A cos\Theta_{Bnd})$ (Table 2, right). The small values of $M_A=$ 0.26 and 0.2 indicate that D1W is a weak SS.
We note that the local flow shear masking the SS-like discontinuity (D1A) has a much larger amplitude than the flow shear  observed by WIND  at the  weak SS (D1W). This is the reason why the SS could be observed by WIND.


Now we consider the inflow region after the right boundary of the outflow (location of $\mathbf{B2}$ vector in Figure 6) which is inside the boundary layer of the MC. The variables in upstream and downstream regions across D2A and D2W are shown in Table 3 left and right.

Across D2A $|\mathbf{V}|$ decreases, the largest changes are in $V_Z$ and $V_Y$ (Figures 7c, d and Table 3, left). Ti increases by 0.2 eV only. There is a  density dip between D2A and the right boundary of the MR outflow. Instead of the sudden Ni jump as it was across D1A, the change from 4 to 10 cm$^{-3}$  is rather smooth across D2A (Figure 7e). In the $\pm$ 10 minute vicinity Ni changes from 7 to 9 cm$^{-3}$ (Table 3, left). Since Ni increases by 1.3-2.5 times  D2A is not a rotational discontinuity in the low beta plasma. Since the normal component of the magnetic field is $\mathbf{B_n}\sim 15.9$, and $\Theta{Bn}$=8.9$^\circ$ and 1.8$^\circ$, respectively,  D2A is not a tangential discontinuity. D2A is associated with a similar velocity stream interaction region as D1A with an embedded quasi-parallel SS-like discontinuity. Therefore, the shock speed cannot be reliably estimated.

Across D2W  $\mathbf{|B|}$ slightly decreases, Ni increases 2.5 times, Ti increases by 0.1 eV, $\mathbf{|V|}$ decreases, however, again the transverse flow speeds $V_Y$ and $V_Z$ show a flow shear at the discontinuity (Figures 7g-k and Table 3, right). $\Theta_{Bn}$ is 14$^\circ$ and 6$^\circ$, respectively. We classify D2W as a quasi-parallel SS-like discontinuity embedded into a flow stream interface.

\subsubsection{MR outflow structure - MHD description}
In the following 
the MR outflow structure is studied in more detail.

Figure 8 shows WIND observations of MR outflow (red vertical dashed lines at 16:46:15 and 16:58:12 UT) in a 1-hour long time interval (between 16.25 and 17:25 UT) in the MVA rotated system (Figure 6c) on 2000-10-03. The discontinuities D1W and D2W are indicated by dashed yellow vertical lines and marked as SS. Although D2W, as described above, is more an SS-like discontinuity in a flow stream, it is also marked as SS.
The top three subplots 8a-c contain the pitch-angle (e-Pa) distribution of 255 eV electrons, the omni-directional electron fluxes for energies from 27 to 520 keV, the omni-directional proton/ion fluxes for energies from 0.4 to 3 keV. Figures 8 d,e show the amplitude and the LMN components of the magnetic field. The ion density and temperature (Ni and Ti) are in the same subplot 8f with the same Y axis in $cm^{-3}$ and eV, respectively. The GSE Vx, Vy and Vz ion velocity components are in Figures 8g and h. With the average ambient solar wind speed removed the LMN ion velocity components are shown in Figure 8i. Additionally, the perpendicular and parallel LMN directional velocity components to the local magnetic field are calculated (Figures 8j,k). The local proton Alfv\'en velocity is also added to the subplot 8j.  Finally, in the last subplot 8l, the on-board calculated electron velocity magnitude $|Ve|$ is shown without the strongly fluctuating components. The LMN components of the low time resolution (one vector/98 s) ground calculated electron velocity are shown in the same subplot 8l. From the electron velocities the average ambient electron speed was also removed.

The MR outflow is as expected in fluid models: simultaneous correlated/anticorrelated rotations of magnetic field and velocity components at the borders (at red dashed vertical lines, Figures 8e,i), inside the outflow decreased $|B|$ (Figure 8d), increased  Ti and Ni (Figure 8f),
and enhanced $\mathbf{L}$  directional velocity $V_L$ (Figure 8i). The component of the velocity perpendicular to the magnetic field $V_{\perp L}$ is almost reaching the local Alfv\'en speed $V_A$ (Figure 8j).
In papers by \citet{wang16} and by \citet{zhao19} the so-called Wal\'{e}n test  has been successfully verified for this event, which is a fluid level test for tangential stress balance for a rotational discontinuity across a reconnecting current layer.
Across the outflow boundaries (red dashed vertical lines) both $\mathbf{V_i}$ and $\mathbf{B}$ rotate, additionally $\mathbf{V_i}$, $\mathbf{V_e}$ and $T_i$ increase towards the outflow center (Figures 8e-h, l). Also, $N_i$ and $N_e$ (see below) clearly increase at the right outflow boundary (towards the center of the outflow at 16:58:12 UT). Such simultaneous  jumps and rotations can correspond to a merged Alfv\'en discontinuity and slow-shock \citep{sasunov12}.
However, the exact jump conditions and discontinuity normal directions cannot be determined because of the high variability of plasma and field data.
Since there are merged Alfv\'en discontinuities and SSs at the boundaries of MR outflow, it is not obvious that the discontinuities D1W and D2W in the inflow regions are Petschek-type SSs. In fact, by using a  compressible resistive MHD model with sub-Aflv\'enic shear flow across the MR outflow it was shown that for low-$\beta$ plasmas a pair of slow shocks (SSs) is generated in the inflow region away from separatrices \citep{li12}.  These SSs in the inflow region in simulations are not the Petschek-type shocks at the separatrices.
In what follows we will refer to the standalone SSs (D1W and D2W in Figure 7) in the inflow regions indicated by the yellow vertical dashed lines  as left SS (at 16:34:33) and right SS (at 17:17:42 UT), respectively. Here left and right just correspond to the time sequence of events, not the spatial locations.

Before continuing with a more detailed description of the outflow structure we shortly describe, without showing, the parameters as they would be in the MVA system (Figure 6b). The L components of variables would not change. The M and N components would be exchanged. For example, ${V_\perp}_N$ and $V_{||N}$ (Figures 8j and k) would be exchanged by ${V_\perp}_M$ and $V_{||M}$. It is easy to notice that, in the MVA rotated system (Figure 6c and the whole Figure 8), at the left and right boundaries of the outflow, the $\mathbf{N}$  directional parallel (minus then plus) and perpendicular ion speeds (in green) are enhanced  and there is a small positive $V_{||L}$ (Figures 8j and k). In fact, these are the MR inflow ion signatures near separatrix when $B_N$ is large (Figure 8e). Similar  inflow signatures are seen also in simulations \citep[e.g.][]{lapenta15}. However, having the MVA coordinates (Figure 6b), the enhanced ${V_\perp}_M$ and $V_{||M}$ components at the outflow boundary would be nonphysical. This seems to support the MVA rotated configuration (Figure 6c.1).
We add that by rotating the coordinate system around L in smaller steps (eg. by 10$^{\circ}$) from the vertical geometry (Figure 6c.1) in clockwise direction Bg would increase in negative direction while in case of counter-clockwise rotation Bg would increase in positive direction.

The right boundary of the outflow at 16:58:12 UT is attached to the MC boundary (time interval indicated by horizontal arrow line, marked as MC over the top panel) which can be readily seen from the appearance of bi-directional suprathermal electrons of 255 eV (e-Pa Figure 8a). The same MC boundary can also be seen in ACE e-Pa 272 eV distributions in Figure 2g. Earlier in time , before the MC boundary, predominantly  the parallel electron population is seen on open field lines, with a smaller flux of anti-parallel electrons starting at the left SS (Figure 8a).
The closed field lines in MC, rooted on the Sun, carry 0.5 MeV low-flux population of electrons of solar origin (Figure 8b). There is also an enhanced electron flux over 30-120 keV between 17:00 and 17:04 UT (Figure 8b). The ion energy spectrogram (Figure 8c) shows two population of ions, one centered at $\sim$950 eV and another centered at $\sim$1900 eV. Inside the MR outflow, before 16:50 UT, both populations are accelerated $\sim$1.4 times. This indicates that both populations of ions are protons. The higher energy proton population appears downstream of the left SS (Figure 8c) where also the low flux anti-parallel electron population is generated (Figure 8a). Similar intensification of the proton flux is seen downstream of the right SS. It is easy to notice that near the left and right quasi-parallel SS (D1W and D2W in Figure 7 and in Tables 2 and 3) there are enhanced fluctuations in $\mathbf{B}$ components (Figure 8e). Magnetic turbulence or waves at quasi-parallel SSs can be responsible for particle acceleration or/and reflection.
The more energetic protons, possibly on the freshly reconnected open field lines,  can propagate from their source within the MR outflow until the left SS where they can be back reflected. Alternatively or concurrently, the reflected protons can propagate back to the outflow where they are further accelerated by MR electric field together with the low energy proton population. The pitch angle distribution of protons in the keV range is not available.
At the right SS a similar increase in the flux of protons/ions in the downstream region is observed, however, the source of the seed population of particles is not clear.

More importantly, there is an asymmetry at the left and right boundaries of the MR outflow including the vicinity where the discontinuities are observed (at $\mathbf{B1}$ and $\mathbf{B2}$ vectors in Figure 6). The left side is located in the ICME sheath, while the right side is in the boundary layer of MC.
Let's consider the left side of the outflow first.

The left SS (D1W above) is associated with a pile-up of plasma and reflection of protons back towards the outflow. Here Ni is increased between the left SS and the left boundary of the outflow.

Both ACE and WIND observe increased Ni after the left SS with slowly decreasing values towards the left outflow boundary (Figures 7c, g). As a result, along WIND crossing, $N_r$=$<$Ni(outflow)$>$/$<$Ni(inflow)$>$ $\sim$ 1.1 near the left boundary of the outflow, where $<$Ni(inflow)$>$ is the average ion density in the inflow region and $<$Ni(outflow)$>$ is the average ion density inside the outflow near its left boundary. Possibly, as the left SS is getting farther or closer to the outflow border, $<$Ni(outflow)$>$/$<$Ni(inflow)$>$ can change and have different impact on the outflow. Following this line of thought, a bundle of freshly reconnected field lines can form a flux tube or flux rope \citep{semenov04, sasunov12, voros14}.
When we consider a bundle to form a twisted flux rope embedded into ambient plasma with a small external twist, Kelvin-Helmholtz (KH) instability can develop even for sub-Alfv\'enic boundary flows \citep{zaqarashvili14}. The criterion for KH instability in such a case is
$ |m|M_A^2 > 1 + N_r(2+|m|)/|m|$,
where $m$ is the azimuthal mode number of KH vortices and $M_A$ is the Alfv\'en Mach number inside the flux rope/outflow. In this context, $|m|$ is the number of vortices which can develop around a flux rope. The analytical and numerical calculations leading to the above instability criterion start from cylindrical flux ropes moving along their axis in an ambient plasma with and without external magnetic twist. Without the external twist, e.g. when the flux ropes move along the ambient magnetic field, that external field appears as an axial one, the KH instability is suppressed for sub-Alfv\'enic flow shear. In our case the internal field inside the current sheet rotates and the external ambient field has a different direction than the $\mathbf{L}$  directional outflow (Figures 8e, i). In such a case the external field has a transverse component and KH instability can develop for a sub-Alfv\'enic flow shear \citep{zaqarashvili14}.
Since $M_A\sim 0.7$ at the left boundary of the outflow and Nr$\sim$1.1, for azimuthal mode numbers $|m|$=5  the KH instability criterion is fulfilled. In fact, between 16:43 and 16:46 UT  low-frequency fluctuations in magnetic field (Figure 8e) and plasma parameters (Figures 8f, h-j) are observed. Although the size and the geometry of the flux rope bundle is not known (probably the flux rope is not a cylinder), the fluctuations clearly seen  in Ti and Ni (Figure 8f)  1-3 minutes before the left outflow boundary could be associated with KH vortices for the sub-Alfv\'enic flow shear.
More importantly, the left SS generating the elevated  density Ni(inflow) and therefore smaller ratio $N_r$ can result in KH unstable outflow boundary with smaller $|m|$ (larger size structures). This can modify the outflow structure through mixing the higher Ni and Ti outflow plasma with the lower Ti and Ni sheath plasma. This is not the case at the right outflow boundary where $N_r\sim$3 (Figure 8f).

Let us consider now the right side of the outflow.
The right SS-like discontinuity (D2W above) with more pronounced transverse flow shear (Figure 8h)  seems to reflect ions (Figure 8c) and increase density downstream of SS (Figure 8f). However, there is a more pronounced density cavity between the right boundary of the outflow and right SS (Figure 8f). The low Ni can appear as a result of pressure balance when MR outflow interacts with the ambient magnetic field compressing it. A similar density cavity has already been observed in magnetotail reconnection associated with interactions between MR outflow and ambient magnetic field in the magnetotail plasma sheet \citep{cai09}. In our case, the cavity after the right boundary of MR outflow is observed by both WIND (Figure 7k) and ACE (Figure 7e). WIND observes a longer low density region, Ni around 5cm$^{-3}$, roughly 20 minutes. The long duration density cavity might also been explained by the slow motion of WIND across the structure. Although $\mathbf{|B|}$ slowly increases in the density cavity, it is unlikely that this can alone generate the observed density cavity due to pressure balance.
When the current sheet configuration is vertical (Figure 6c) the right boundary of the outflow is slowly convected in +Z GSE  direction with $V_Z\sim$ 6 km/s  (Figure 7j). The fast motion with the bulk solar wind speed in GSE-XY directions moves the vertical structure mainly in the L-M plane (Figure 6c.1), therefore, WIND possibly remains near the separatrix surface where the low Ni is observed. We note the separatrix is a line in 2D geometry while it is a surface in 3D case.

\subsubsection{MR outflow structure - kinetic description}
At the outflow boundaries the  high-frequency field and plasma fluctuations are enhanced. Although in MR separatrix parallel electric fields, waves, instabilities and kinetic physics are expected \citep{khotyaintsev20, hesse18}, the limited time resolution of field and plasma measurements makes the investigation of sub-ion and electron scale structures more difficult. Nevertheless, we demonstrate here that WIND can observe signatures of electron-scale physics in MR outflow.

First, we concentrate on the narrow structure occurring inside the outflow at 16:48:39 UT inside the magenta dashed box in Figure 8. At and around that time, the accelerated protons reach their maximum energy (Figure 8c); $B_L$ changes sign sharply decreasing from 5 to -12 nT, $B_N$ decreases sharply from 12 to 6 nT, $B_M$ changes from -6.5 to 0 nT; the guide field Bg $\sim$ -2 nT  (Figure 8e); Ni decreases and Ti increases (Figure 8f); $V_{\perp N}$ which is predominantly negative within the outflow locally reaches +40 km/s (Figure 8j); $V_{|| N}$ reaches locally -70 km/s, $V_{|| L}$ +100 km/s (Figure 8k) and the magnitude of onboard calculated electron velocity $|V_e| \sim 220$km/s which is by 130 km/s larger than the ambient electron outflow velocity (Figure 8l). The lower resolution ground-calculated  ${V_e}_{L}$ has the same peak as the onboard computed $|V_e|$.
The accelerated protons, the increased parallel velocities
$V_{|| L,N}$ and increased $|V_e|, {V_e}_L$ in Figure 8 indicate that  there might be a local parallel electric field $E_{||}$ which accelerates particles and therefore generates currents. In principle, the current density could be determined from the curl of $\mathbf{B}$ through $J_L \sim -\partial B_M/\partial N  +  \partial B_N/\partial M$, however, the field and plasma parameters, including the Hall-field ($B_M-Bg$) fluctuate wildly. Turbulent Hall fields associated with MR have already been observed in the solar wind \citep{xu15}. Qualitatively, the currents responsible for the Hall field can be generated by electrons flowing along the separatrices, ie. near the outflow border, and at the narrow current structure (in dashed magenta box) where proton $V_{||L}$ (Figure 8k) and $Ve_L$ (Figure 8l) are large. The latter narrow current structure occurs closer to the left border of the outflow (closer to $\mathbf{B1}$ in Figures 6b or 6c). The deflection of the $J_L$ current from the midplane of the outflow has been reported in simulations, \citep{goldman11}, far from the X-line in the solar wind \citep{xu15, mistry16} and in magnetotail reconnection \citep{zhou14}. For $Ve_L > V_{||L} >$0 the current is in negative $\mathbf{L}$  direction, pointing towards the X-line. The deflection of electrons is supposed to occur due to the Lorentz force in the presence of a weak guide field given by -q.$Ve_L \times Bg$, where q is the electric charge. Since Bg = -2 nT (Figure 8e, Bg/$|\mathbf{B}|\sim$ 0.13) and  $Ve_L$ is positive (in the magenta dashed box, Figure 8i), the Lorentz force deflects the electron jet towards the left boundary of the outflow. For the vertically oriented current sheet with Bg in $\mathbf{-M}$  direction (Figure 6c.1) it is the boundary at $\mathbf{B1}$. However, it is also valid for the MVA LMN coordinates when LN is in GSE-XY plane in Figure 6b. In that case the large positive guide field Bg$\sim$7.5 nT (Figure 5b, Bg/$|\mathbf{B}|\sim$ 0.5) ensures that the Lorentz force deflects the electron jet again towards $\mathbf{B1}$.

The current layers associated with deflected electron flows are expected to be narrow of 0.8-6 di \citep{mistry16}. di stands for ion inertial length. Since di$\sim$ 100 km in our case, when the LN plane is in GSE-XY (Figure 6b), the thin electron flow structure should be convected with $V_{SW}$ velocity in less than 2 s across WIND. WIND does not have a good enough temporal resolution to see such a thin electron flow. Nevertheless, it is  observed, as shown within the dashed magenta box in Figure 8l. However, in the MVA rotated system the LM plane is in the GSE-XY plane (Figure 6c) and the vertical convection speed of the structure is $\sim$16 km/s (Figure 7j) or less. With this speed the electron flow structure could be observed by WIND for more than 30 s, provided that the flow channel is more extended along the X-line, in out-of-plane direction. It is worth noticing that simultaneously with the occurrence of the narrow electron and ion flows in the dashed magenta box, $B_L$ transiently changes sign and $B_N$ decreases (Figure 8e). This might be the signature that the thin flow structures locally change the magnetic field geometry inside the outflow. Interestingly, Geotail, which is at a larger distance from the X-line as WIND, observes the sign-changing $B_L$ roughly 10 s earlier (Figures 5d, e). If we accept that the transient short sign-change of $B_L$ is the signature of the interaction of the flow with the magnetic field then the electron flow is possibly even more  deflected towards the left border of the outflow at Geotail location, which is in a larger distance from the X-line.

There are other kinetic effects observed by WIND which can also  shape the MR outflow structure. The right boundary of the outflow was previously classified as a merged slow shock and Alfv\'en discontinuity at the MHD level of description. Now we consider possible kinetic effects at the same boundary.

In Figure 9a  the MVA rotated LMN components of $B$ and its magnitude are shown. Figure 9b contains Ni and Ne, the latter  estimated from the WAVES experiment  (based on electron plasma frequency) on WIND. Ti and Te are shown in Figure 9c. The electron temperature anisotropy $Te_{||} / Te_{\perp}$ is depicted in Figure 9d. Figure 9e shows again the omnidirectional electron energies in keV-MeV range and in Figures 9f-j are the pitch angle distributions (e-PA) for energy ranges from 40 to 310 keV. The frequency spectra of the average electric field in dB above the background and the normalized average voltage from the Thermal Noise Receiver (TNR) experiment on Wind/WAVES are shown in Figures 9k and l, respectively. The outflow is again indicated by red vertical dashed lines.

The magenta dashed box in Figure 9 indicates the time interval between the right end of the outflow and right SS. At the beginning of this time interval the  electric field fluctuations are enhanced and wave activity is present in Figures 9k and l. This is the time interval where both Ni and Ne are low and, as it was explained above, the outflow interacts with the ambient plasma and magnetic field of the MC. While Ti is increased within the outflow, Te is enhanced in the density cavity (Figure 9c) where there is also a clear temperature anisotropy
$Te_{||} / Te_{\perp} \sim$1.5 (Figure 9d). This indicates that electron heating is preferentially in parallel (or anti-parallel) direction to the magnetic field.
In the density cavity the e-PA distributions for 40 and 86 keV energy range show perpendicular populations of electrons (Figures 9f,g). The local enhanced flux over this energy range is also clearly seen in the omnidirectional electron energy spectrum (Figure 9e). Over the 180-310 keV energy range the electrons are predominantly anti-parallel (Figures 9i,j) with a smaller flux in the parallel direction. There is also an isotropic electron background population with smaller flux which is already present inside the outflow near its right boundary at higher electron energies (Figures 9i, j). For the 180 - 310 keV electrons the e-PA is bi-directional indicating the presence of closed field lines in the MC. We have also checked a 3-hour long time interval after the outflow  and bidirectional e-Pa distributions associated with MC closed field lines are present for the energy range of 110-520 keV (not shown). The strongest fluxes of anti-parallel electrons occur in the density cavity.
We have also checked the lower energy electron e-PA distributions in the wider time interval.
Although there exist  perpendicular or more isotropic electron populations for energies 27-86 keV even 3 hours later (not shown), the perpendicular electron population in the density cavity close to the right outflow boundary (Figures 9f, g) has the largest flux.
Also, the density cavity is the only place during the 3 hour long wider time interval (not shown), where the electric field fluctuations peak around 10 kHz (Figure 9k) and where wave activity is observed over the frequency range of 5-20 kHz (Figure 9l).

Now we provide a possible physical explanation for e-Pa distributions and  wave activity seen in the density cavity and before, inside the outflow  (Figure 9).

Let us start with the observations of electron acceleration in the Earth's magnetosphere.
For example, in the Earth's magnetotail electrons accelerated up to $\sim$300 keV have been observed during a crossing of MR diffusion region \citep{oieroset02}. However, electron acceleration by MR in the magnetotail can be a multi-stage process, when electrons are accelerated to a few keV in the vicinity of the X-line by MR electric field, and further nonadiabatically accelerated to tens of keV in the outflow pile-up region  due to $\mathbf{\nabla } B$ or/and curvature drift where the electron flux is also substantially increased \citep{wu15}.

In our case, from one-point measurements the magnetic field curvature or $\mathbf{\nabla } B$ cannot be estimated.
We provide a qualitative description of $\mathbf{\nabla } B$-drift  and a rough estimate of the spatial scales associated with a possible  electron acceleration.

Near the right boundary of the outflow $\mathbf{|B|}$ increases from 11 to 14 nT in $\sim$35 s, which is mainly due to an increase of $\mathbf{|B_L|}$ from 5 to 14 nT  (Figures 8 e,f). When the current sheet normal is vertical (Figure 6c) and the right boundary of the current sheet moves slowly in vertical direction ($V_Z\sim$6$\pm$6 km/s in Figure 7j), the $\sim$35 s time interval corresponds roughly to 200$\pm$200 km spatial scale. Since the gyroradius of 40-310 keV relativistic electrons is between 50 and 200 km, comparable to the spatial scale of $\mathbf{\nabla } B$, nonadiabatic chaotic motion of electrons can be expected \citep{wu15}. This can be readily seen in the e-PA spectra of 180-310 keV electrons of larger gyroradius (Figures 9i, j), where an isotropic background distribution, possibly due to the chaotic motion, is visible near the right end of the outflow.

When $\mathbf{|B_L|}$ increases by 9 nT at the right outflow boundary, GSE $\mathbf{|V_X|}$ decreases from 430 to 390 km/s (Figure 8g),  GSE $\mathbf{V_Y}$ decreases from 50 to -5 km/s  and GSE $\mathbf{V_Z}$ changes from -20 to 20 km/s (Figure 8h). Although the rotation of the velocity direction is an indication for the Alfv\'en discontinuity, however, $\mathbf{|V|}$ locally decreases by $\sim$35 km/s. This indicates that when the outflow interacts with the ambient plasma and magnetic field of the MC, the plasma inside the outflow near the right outflow boundary decelerates and the magnetic field is compressed. The largest compression occurs near the right outflow boundary, but $\mathbf{|B|}$ continues increasing roughly until the right SS (Figure 8d), within the whole density cavity.
As the kinetic energy of the plasma decreases  particles are energized at $\mathbf{\nabla } B$ structure.  Due to the changing velocities it is not possible to estimate precisely $\mathbf{\nabla } B$ during this time interval from single-point measurements. In case of the vertically oriented current sheet (Figure 6c) $|B_L|$ ($= \mathbf{B_2}$) is increasing in $\mathbf{-N}$ direction. Due to $\mathbf{\nabla } B$-drift the electrons move into $\mathbf{+M}$ out-of-plane direction. In the density cavity the magnetic compression is smaller, however,    $\nabla B$ - drift can still generate a current and the electrons and protons move in opposite but perpendicular directions to $\mathbf{B}$ and $\mathbf{\nabla } B$. This can explain the occurrence of the perpendicular population of accelerated electrons  near the outflow right boundary in Figures 9f and g. However, the largest flux of perpendicular electron populations is seen 2-6 minutes after the right outflow border or the strong $\mathbf{\nabla } B$ structure (Figure 9d,e).
The question is what is the real spatial distance between the location of the right outflow boundary and the location of perpendicular electrons (2-6 minutes later in time) or the right SS. How wide is the magenta dashed box in space with $\sim$20 minutes duration in time? Here again, the geometry of the current sheet matters (Figure 6b versus Figure 6c).
When the outflow structure is convected vertically in GSE Y-Z direction (case Figure 6c)  WIND can remain close to the right flow boundary  or near the 2D separatrix for longer time. The average GSE $V_Z$ near the right outflow boundary is 6 km/s, however, fluctuating (Figure 7j).
Within 2-6 minutes the outflow border / separatrix can move to the distance of 2000$\pm$2000 km in vertical direction, within 20 minutes the width of the magenta dashed box is 7000$\pm$7000 km. These are certainly rather rough estimates of the spatial distances.
We note that for the geometry in Figure 6b, the distance of the location of perpendicular electron population to separatrix (where $\mathbf{\nabla } B$ is significant)  would be 50000-100000 km (8-16 $R_E$) and the width of the magenta box would be 516000 km (81 $R_E$).

We suggest here that both the perpendicular and the anti-parallel relativistic electron populations can be the result of a multi-stage acceleration by MR. In the vertical configuration (Figure 6c) the 20 minute long magenta dashed box can coincide with the reconnection inflow region, spatially close to the 2D separartix. While the lower energy electrons react to the $\mathbf{\nabla } B$  structure which is strong near the right outflow boundary, the higher energy electrons are anti-parallel, which at the right outflow boundary (at $\mathbf{B2}$ in Figure 6c) means propagation from the X-line.
Although the details of multi-stage electron acceleration are not clear,
the enhanced electric field fluctuations and wave activity in Figures 9i and j can be associated with the relativistic electrons.

The WIND/WAVES experiment allows to detect high-frequency plasma waves between local ion (fpi) and electron (fpe) plasma frequencies. In this frequency range, by using the TNR spectra and the electric waveforms from the Time Domain Sampler (TDS), \citet{huttunen07} observed Langmuir waves, electron solitary waves and Doppler shifted ion acoustic waves (around $\sim$4 kHz) inside the MR outflows,  preferentially near the boundaries of MR outflows in the solar wind. Since TDS waveforms are not available for the event under study we cannot distinguish between Langmuir and upper hybrid waves, both occurring near fpe  which is about $\sim$20 kHz in the magenta dashed box in Figure 9j. The waves observed below fpe down to 5 kHz in TNR spectra are supposedly  Doppler shifted ion-acoustic waves \citep{huttunen07}. Lower frequency waves cannot be detected by the WIND/WAVES experiment. According to 2D particle-in-cell simulations of antiparallel MR \citep{fujimoto14} the Langmuir waves can be generated by intense parallel electron beams triggering the electron two-stream instability. In these simulations an electrostatic potential is created due to electron-ion decoupling near the separatrix - inflow region forming the Hall current system supported mainly by electron beams. In the potential well the electrons are accelerated and the density decreases to conserve the mass flux. As a result, a density cavity can be formed in the inflow/separatrix region \citep{fujimoto14}. In  the simulation the 2D box was 131 $\times$ 65.5 di only, i.e. the distant MR outflow and separatrix/inflow regions were not studied. In our case however, there exist Hall-field signatures. The antiparallel electrons in the right separatrix/inflow region (at $\mathbf{B2}$ in Figure 6c) indicate a current flowing towards the X-line which should generate a positive Hall magnetic field in the inner part of the right outflow boundary. Indeed, Figure 9a (or Figure 8e) demonstrates that
relative to the small Bg, B$_M$ is fluctuating, but positive. As it was noticed above, it was unlikely that the increased $\mathbf{B}$ and the pressure balance could explain the occurrence of the density cavity in the inflow region. In the presence of the Hall current system the subsequent acceleration of electrons in the potential well and the requirement for mass flux conservation can explain the generation of the density cavity.
Alternatively, the electron beams can trigger the Buneman instability resulting in waves with frequencies near the lower hybrid range \citep{fujimoto14} (out of the range of WIND/WAVE). The Buneman instability is expected to be excited for stronger guide fields \citep{drake03}.

\section{Summary and conclusions}

In this paper MR outflow structure at the boundary layer of the MC occurring between 2000-10-03 and 2000-10-04, associated with an ICME, was  studied. The ICME was considered in a wider time interval, between 2000-10-02 and 2000-10-05. First, the orientation of the MC axis, MC helical structure and size, the width of the ICME sheath was determined (Figures 1-4). Multi-point measurements from ACE, Geotail and WIND (Figure 1) have been used whenever it was possible. Although at or near the trailing edge of the MC there exist interactions with shocks and fast ejecta (Figure 2), in the immediate vicinity of MR outflow  interactions with shocks are absent. This means that there are no fast interplanetary shock-induced local interactions, particle accelerations or field deformations which could make the observation of MR outflow structures more complicated.

The main findings and some additional speculations are summarized in Figure 10. The cartoon in Figure 10a shows the large-scale ICME, the left handed helical structure of the MC with axis pointing towards the Sun. The GSE XYZ directions are shown on the top left part of the subplot. Inside the MC two partial helical field lines are shown only (enough to demonstrate the main ideas), one closed field line (brown)  and another one (black) which is the freshly reconnected field line. The field lines over the MC axis are depicted by thick lines, below the axis thin lines are used. Half turning of the magnetic field around the axis has a spatial dimension of approximately 5000 R$_E$. The reconnection X-line at the western flank of the structure is between the observed reconnection jet  and the Sun. The jet moving in +L direction is subsequently observed by ACE, Geotail and WIND, while SOHO does not observe any outflow signatures. The distance from the spacecraft outflow observations to the X-line is not known, it is, however, $\gg$ 314 $R_E$. The outflow-pair is indicated as two "jets", one propagating towards the Sun associated with closed field lines and one propagating towards the spacecraft associated with open freshly reconnected field lines. The open field lines enable plasma mixing resulting in higher O$^{7+}$/O$^{6+}$ ratios than the background values (Figure 2h) and intermediate plasma temperatures between the sheath and MC. We call the plasma mixing region as mixing layer (ML, Figure 2) which is indicated by gray shadow sprayed area in the sheath in Figure 10a.
In the sheath there are also open field lines emanating directly from the Sun (i.e. not through the MC, magenta line). These field lines carry parallel electron populations, which can be seen in Figure 8a, before the MR outflow. Near the jet propagating towards the spacecraft the reconnection vertical LMN system is shown (Figure 6.c.1). It can be seen in Figure 10a that the $\mathbf{L}$  direction is roughly antiparallel to the MC axis. The width (L1) of the sheath along SOHO, ACE or Geotail trajectories  is shorter than along WIND trajectory (width: L2), this is an indication that the structure is crossed at its western flank. For simplicity two trajectories (red vertical dashed lines) are shown only.  For the left handed helical structure,  MC axis pointing towards the Sun, IMF pointing in anti-sunward direction, and MR outflow in roughly anti-sunward direction, the reconnecting MC magnetic field lines reach the X-line from below of the MC axis. This means that part of the freshly reconnected open field lines are wired to the MC while the other part of the field lines freely moves in the turbulent sheath region.
In the sheath, the newly reconnected open field lines are interacting with the ambient plasma. It can be seen in the event overview plot (Figure 2) that 4 hours downstream of the MC driven shock S1 the $V_Z$ GSE velocity is reaching $\sim$ 60 km/s (Figure 2e). Within the mixing layer (blue dashed box) $V_Z$ starts decreasing until it reaches 18 km/s before the left boundary of MR outflow (also, Figure 7b). $B_Z$ associated with $V_Z \sim$60 km/s in the sheath fluctuates around 0 nT (Figure 2c). Within the mixing layer there are also newly reconnected field lines having low $V_Z$ velocity close to MR and  higher $V_Z$ velocity field lines within the mixing layer. Since the magnetic field is frozen into the plasma flow, as $V_Z$  decreases towards MR outflow, $B_Z$ becomes positive (Figures 2c, e). This might indicate that the newly reconnected field lines interacting with plasma in the sheath slip over the MC in +Z GSE direction (not shown in Figure 10c). In this way,for the given MC and MR geometries and plasma flows, northward oriented (+B$_Z$) magnetic field  is generated in the ML which is wired to the helical MC (field lines  below MC axis). In the opposite case, when the field lines were slipping below the MC the newly reconnected field lines, for the given geometry, would disappear more easily from the sheath and perhaps the ML would not be formed (not shown in Figure 10a). It is also easy to recognize, that in a case when the MC axis was pointing towards the Sun but with right-handed  helicity, IMF in the sheath was oriented from the Sun, and magnetic field lines were slipping below the MC, a geoeffective -B$_Z$ would be generated within the mixing layer in the sheath (not shown). In the ICME sheaths there also exist other processes which can influence the geometry of the magnetic field in front of the MCs. We mention here the pile-up of the solar wind ambient plasma in front of the MC when the sheath is expansion type \citep{siscoe08}. In our case the relative velocity between the leading edge of MC and upstream ambient solar wind is $\sim$ 30 km/s while the MC expansion speed is $\sim$25 km/s. It indicates that the sheath is of  hybrid type, i.e. a mixture of expansion-propagation sheath, in which the  lateral deflection of plasma away from the nose of the MC is slow \citep{siscoe08}. In fact, in the mixing layer containing also  the reconnected field lines the transverse velocities (Figure 2e) are of the order of the MC expansion speed, indicating that the slipping of the field lines over the obstacle (MC) is not too fast.

Figure 10b and c show the cartoons of the MR outflow structure reconstructed from data analysis. For the sake of better visualisation, we have chosen to show two cartoons (quasi-3D and the 2D LN plane) of the same outflow. The quasi-3D cartoon is the spatial segment containing the X-line, the outflow crossing (one trajectory is shown only) extending from the ICME sheath to the MC boundary layer.

The outflow vertical geometry in Figures 10b,c correspond to Figure 6.c.1, where N $\sim$ Z GSE and the LM directions lay in XY GSE plane. The GSE and LMN coordinates are shown on bottom right side of Figure 10b. $B_N$ is positive and the average vectors $B_1$ and $B_2$ are shown according to the notation in Figure 6.c.1. As it was discussed above several times, the outflow at the boundary layer of the MC is embedded into a large-scale $V_Z$ flow shear (Figure 2). In the vicinity and across the outflow the positive $V_Z$ in the sheath changes to zero and then to negative $V_Z$ in the MC. At WIND, near $B_1$ $V_Z \sim$ 16 km/s near $B_2$ $V_Z = 6\pm6$ km/s (Figure 7 and Figure 10b). This shows that the whole outflow structure moves in +Z GSE direction and this motion decelerates to a small speed near $B_2$. This is also indicated in Figure 10c. WIND trajectory across the outflow (from $B_1$ to $B_2$) is shown by a thick magenta line in Figure 10b and magenta dashed line in Figure 10c. Other motions in XY-GSE or LM directions move the outflow structure in an N$\sim$const. plane. It was found that in $\mathbf{M}$  direction the outflow is roughly 3050 di wide, while the thickness of the structure in $\mathbf{N}$  direction is about 157 di (Figure 10b).

Using multi-spacecraft triangulation and the local $V_Z$ velocities observed by the spacecraft it was also shown that the vertically moving outflow can be observed by ACE, Geotail and WIND, but not by SOHO, which is, in fact, the case here.

In addition to the merged rotational discontinuity and SS at the boundaries of the outflow (shown by beige color in Figure 10c), in the inflow regions SS-like discontinuities (D1W and D2W in Tables 2 and 3) associated with local flow shears were observed mainly by ACE and WIND (Figure 7). In the sheath inflow  region (at $B_1$ in Figures 10b,c) the SS-like discontinuity observed by WIND  tends to increase the ambient density towards the outflow border to almost the level observed in the outflow (Figure 8f). As a consequence, Kelvin-Helmholtz instability could generate local fluctuations even for the sub-Alfv\'enic outflow shear. The normal vectors of D1W and D2W (Tables 2 and 3) transformed from GSE to LMN coordinates are $\mathbf{n1}=[0.9\;\; -0.2\;\; 0.42]$ and \hfill \break $\mathbf{n2}=[0.77\;\; 0.16\;\; -0.55]$, indicated by blue arrows in Figure 10c). Interestingly, the roughly V-shaped pattern behind the moving obstacle (outflow) resembles  the V-shaped wake waves behind moving ships on calm water \citep{rabaud13}.

Despite the limited time resolution of field and plasma data on WIND we were able to observe kinetic effects which can shape the MR outflow structure as well.

The kinetic effect occurring near the sheath boundary of the outflow is the deflection of the electron jet by the Lorentz force -Ne.q.($\mathbf{Ve_L}\times \mathbf{Bg}$) outflow. The deflected electron jet is observed close to the left boundary of the outflow ($\mathbf{B_1}$) shown in Figure 8f and also in Figure 10c. The current generated by the differential motion of fast electrons and slower ions contributes to the Hall field. Nevertheless, at the same time protons are accelerated from $\sim$1 to $\sim$1.4 keV energies (Figure 8c). The protons on open field lines can reach the SS-like left discontinuity, where they are probably accelerated to $\sim$ 2 keV and back scattered to the outflow where they reach $\sim$ 2.8 keV energy (Figure 8c). We note that the magnitude and sign of the Bg field  and $B_{Hall}$ change when the MN directions are rotated around L.

Several other kinetic effects have been observed near the right boundary of the outflow. In the MC inflow region (at $B_2$ in Figures 10b,c) a density cavity is observed which is associated with anti-parallel (from X-line) and perpendicular to $B$ relativistic electrons, Langmuir and Doppler-shifted ion-acoustic waves between 5-20 kHz. Since the high-frequency waveforms are not available for this event, these waves are identified here  on the basis of the frequency ranges, which seems to be also supported by previous statistical results on observations of these waves associated with reconnection outflows in the solar wind \citep{huttunen07}. Nevertheless, the occurrence of the density cavity near the separatrix region suggests that the relativistic electrons are accelerated in the potential well associated with the Hall current system and the cavity is the consequence of mass flux conservation rather than pressure balance. The perpendicular population of electrons near the separatrix in inflow region (at $\mathbf{B_2}$, Figures 9f,g) can be explained by $\Delta B$ drift which drives electrons into $\mathbf{+M}$  direction. Inside the outflow near $B_2$ the compressions imposed by the flow on the ambient field and plasma can be associated with nonadiabatic acceleration of electrons. This is best seen for 180-310 keV scattered electrons (Figure 9i,j), when their gyroradius was found to be comparable to the estimated spatial in-homogeneity scale (not shown in Figure 10).  There exists also relativistic bidirectional electrons on closed MC field lines, probably associated with the source regions on the Sun. For better visibility, the electron propagation directions are shown downstream of $B_2$ along $\mathbf{L}$  direction in Figure 10b. We mention that \citet{wang16} observed also a lower flux of omni-directional protons with a peak at $\sim$100 keV near the right boundary of the outflow.

The alternative for MR LMN system would be LN laying in XY GSE plane and M$\sim$ Z GSE (Figure 6b). In this case the thickness of the current sheet along N would be more than 3000 di. Also, for this current sheet geometry we would have a large guide field, not structured, unipolar and large Hall field, nonphysical $V_M$ velocity components at flow boundaries (see the discussion of Figures 8j,k for this case). Additionally, we believe that the kinetic signatures of Lorentz force deflected electrons  and the relativistic electrons and the associated density cavity would not be observed in this geometry, given the time resolution of WIND and the high bulk velocity of the solar wind.

We note that the vertical geometry of the current sheet as it is presented in Figure 6.c.1 and Figures 10 b,c means that the LN reconnection plane or the $\mathbf{N}$  direction is roughly perpendicular to the normal vector of MC (tangential to the surface of MC). The MC normal vector could not be determined for the event studied here. On the other hand, MCs are usually fitted by cylindrical or toroidal models with smooth surfaces. At the terrestrial magnetopause the reconnection $\mathbf{N}$  direction is expected to be roughly perpendicular to the magnetopause smooth surface  \citep[e.g.][]{burch16}. The local geometry of MC surfaces over spatial scales relevant for 3D reconnection physics (from electron scales to hundreds or thousands of di, up to tenth of  R$_E$) are not known in the solar wind. It is quite reasonable to suppose that, due to the CME associated eruptive source region processes on the Sun and due to the interactions with the non-homogeneous solar wind, there are small- or medium-scale irregularities forming the surface of an MC. As a consequence, the MC boundary formed by twisted magnetic field lines can resemble a braided hair or dreadlocks rather than a smooth cylindrical or toroidal surface. Under such geometry a 3D MR can take place and the $\mathbf{N}$  direction is more arbitrary. By rotating the $\mathbf{M N}$  directions around $\mathbf{L}$  by 10-20 degrees the above outlined physical picture  for the vertical geometry does not change too much. Also, we emphasize here the importance of a careful MVA analysis for determination of the MR LMN  coordinates in nested time intervals in the solar wind. When possible, additional physical tests for the determination of the relevant LMN coordinate system has to be accomplished. In this paper we treated two cases: the customary MVA and the rotated MVA coordinate systems. In the first case, the eigenvalue ratios  of the variance matrix were satisfactorily large in wider time intervals across the outflow. We believe, this is the usual approach for determining the MR coordinate system in the solar wind. Other approaches based on cross products for determination of current sheet normal directions work only when the angle between the corresponding vectors is sufficiently large. In the second case, in nested time intervals of decreasing lengths we have found  MVA degeneracy in perpendicular to $\mathbf{L}$  directions and on this basis we considered transformations of field and plasma parameters using the MVA rotated system. We believe, the relevant coordinate system is the rotated vertical one (Figure 6.c.1) in which the outflow structure can be properly understood (Figure 8).

In summary, we have demonstrated that the MR outflow occurring in the MC boundary layer strongly interacts with the ambient plasma and magnetic field. This is supported by the main results of the paper:
\begin{itemize}
    \item A plasma mixing layer is generated in the ICME sheath where via MR and transverse flows, ahead of the MC, the large-scale magnetic field is re-organized;
    \item the MR outflow is embedded into a large-scale vertical flow shear in the spatial region with small vertical speeds;
    \item The Petschek-type MR outflow additionally generates non-Petschek type SS-like discontinuities in the inflow regions associated with local gradients in the vertical flow shear;
    \item The SS-like discontinuities can have a back effect on the outflow structure, for example, supporting Kelvin-Helmholtz instability and plasma mixing at the outflow border;
    \item The Lorentz-force deflected electron jet can contribute to the appearance of a turbulent Hall field
    in a large distance from the MR X-line;
    \item Protons are accelerated by MR electric fields; presumably the same proton population is back reflected/accelerated  from the SS-like discontinuity in the inflow region along the freshly reconnected field lines;
    \item Near the MC inflow boundary/separatrix  relativistic electron populations scattered, perpendicular and antiparallel to the magnetic field are observed. These electron populations can presumably appear as a result of non-adiabatic acceleration, energy exchanges, $\mathbf{\nabla } B$-drift and via acceleration in the electrostatic potential well associated with the Hall current system. The density cavity at the MC inflow region appears to be  the result of mass flux conservation.
    \item Associated with relativistic electrons, Doppler shifted ion-acoustic and Langmuir waves are observed, the latter supposedly being generated through electron two-stream instability.
\end{itemize}

We believe that the fluid scale structures observed in the MC boundary layer, in the vicinity of MR outflow or the processes associated with MR outflow structure (the first 4 points in the above list) could be routinely observed near or at the boundary of other MCs where MR occurs.  Regarding kinetic effects (the last 4 points in the above list), the specific current sheet geometry and its slow vertical motion, required for WIND observations of small-scale structures, certainly represent fortunate observational circumstances which occur only occasionally in the solar wind. The importance of kinetic physics in solar wind MR events can be confirmed by missions with high resolution particle and field data products, such as Parker Solar Probe or Solar Orbiter.

%


%
%
%
%

\begin{table}
 \caption{MVA coordinate systems in GSE}
 \centering
\begin{tabular}{|l||l|l|}
  \hline
Spacecraft        & MVA LMN eigenvectors &  MVA-rotated LMN eigenvectors \\
\hline \hline
ACE   &    L=[-0.62 0.78 -0.08]    &    L=[-0.62 0.78 -0.08]      \\
      &    M=[0.17 0.03 0.99]    &    M=[-0.77 -0.62 0.15]       \\
      &    N=[0.77 0.62 -0.15]    &    N=[0.17 0.03 0.99]       \\
\hline
Geotail   &    L=[-0.63 0.78 -0.06]    &    L=[-0.63 0.78 -0.06]      \\
      &    M=[0.15 0.02 0.99]    &    M=[-0.77 -0.63 0.11]       \\
      &    N=[0.77 0.63 -0.11]    &    N=[0.15 0.02 0.99]       \\
\hline
WIND   &    L=[-0.65 0.75 -0.07]    &    L=[-0.65 0.75 -0.07]      \\
      &    M=[0.05 0.05 1]    &    M=[-0.76 -0.65 0.01]       \\
      &    N=[0.76 0.65 -0.01]    &    N=[0.05 0.05 1]       \\
\hline
\end{tabular}
\end{table}

\begin{table}
 \caption{Discontinuities in the ICME sheath}
 \centering
\begin{tabular}{|l||l|l|||l||l|l|}
  \hline
\multicolumn{3}{|l|||}{\bf{D1A}:    $ \mathbf{n}=$ [-0.96 -0.24 0.13]} GSE &
\multicolumn{3}{|l|}{\bf{D1W}:      $\mathbf{n}=$ [-0.42 0.83 0.35]} GSE\\
\hline
Variable        & upstream & downstream &  Variable   & upstream & downstream \\ \hline \hline
$\mathbf{|B|}$ [nT]   &    16.5    &    15.8     & $\mathbf{|B|}$ [nT]&    15.4   &    14.5 \\
$B_X$   &    -9.4    &    -8.6     & $B_X$&    -9   &    -7.3 \\
$B_Y$   &    12.9    &    11.5     & $B_Y$&    11   &    11.8 \\
$B_Z$   &    4.3    &    6.7     & $B_Z$&    5.9   &    4.2 \\
\hline
$\mathbf{|V|}$ [km/s]  &    415.8    &    414.3     & $\mathbf{|V|}$ [km/s]&    418   &    426 \\
$V_X$   &    -414    &    -414     & $V_X$&    -417   &    -425 \\
$V_Y$   &    12    &    7     & $V_Y$&    28   &    27.5 \\
$V_Z$   &    37    &    15     & $V_Z$&    14   &    21 \\
\hline
$Ni$ [cm$^{-3}$]  &    8    &    13     & $Ni$ [cm$^{-3}$]&    8.2   &    10.5 \\
\hline
$Ti$ [eV]  &    1.3    &    2     & $Ti$ [eV]&    2.2   &    3 \\
\hline
$\Theta_{Bn}$ [$^{\circ}$]  &    67    &    67     & $\Theta_{Bn}$ [$^{\circ}$] &    12   &    4 \\
\hline \hline
$\mathbf{V}_{sh}$ [km/s]  &  \multicolumn{2}{|l|||}{-}     & $\mathbf{V}_{sh}$ [km/s]&    \multicolumn{2}{|l|}{229}    \\
\hline
$U_n$ [km/s]  &    -    &    -     & $U_n$ [km/s]&    25.7   &    20 \\
\hline
$M_A$  &    -    &    -     & $M_A$ &    0.26   &    0.2 \\
\hline
\end{tabular}
\end{table}

\begin{table}
 \caption{Discontinuities within the boundary layer of MC}
 \centering
\begin{tabular}{|l||l|l|||l||l|l|}
  \hline
\multicolumn{3}{|l|||}{\bf{D2A}:    $ \mathbf{n}=$ [-0.65 0.49 -0.58]} GSE &
\multicolumn{3}{|l|}{\bf{D2W}:      $\mathbf{n}=$ [-0.65 0.45 -0.61]} GSE\\
\hline
Variable        & upstream & downstream &  Variable   & upstream & downstream \\ \hline \hline
$\mathbf{|B|}$ [nT]   &    16.2    &    15.9     & $\mathbf{|B|}$ [nT]&    14.86   &    14.78 \\
$B_X$   &    10.1    &    10     & $B_X$&    10.45  &    8.5 \\
$B_Y$   &    -9.9    &    -7.8     & $B_Y$&    -8.55   &    -6.5 \\
$B_Z$   &    7.9    &    9.6     & $B_Z$&    6.2   &    10.2 \\
\hline
$\mathbf{|V|}$ [km/s]  &    427.8    &    426     & $\mathbf{|V|}$ [km/s]&    436.5   &    434.2 \\
$V_X$   &    -427    &    -426     & $V_X$&    -435   &    -432 \\
$V_Y$   &    15    &    5     & $V_Y$&    35   &    42 \\
$V_Z$   &    21    &    -2     & $V_Z$&    8   &    -11 \\
\hline
$Ni$ [cm$^{-3}$]  &    7    &    9     & $Ni$ [cm$^{-3}$]&    4.5   &    11.2 \\
\hline
$Ti$ [eV]  &    0.9    &    1.1     & $Ti$ [eV]&    2.4   &    2.5 \\
\hline
$\Theta_{Bn}$ [$^{\circ}$]  &    8.9    &    1.8     & $\Theta_{Bn}$ [$^{\circ}$] &    14   &    6 \\
\hline
\end{tabular}
\end{table}

   \begin{figure}[h]
 \includegraphics[width=15cm]{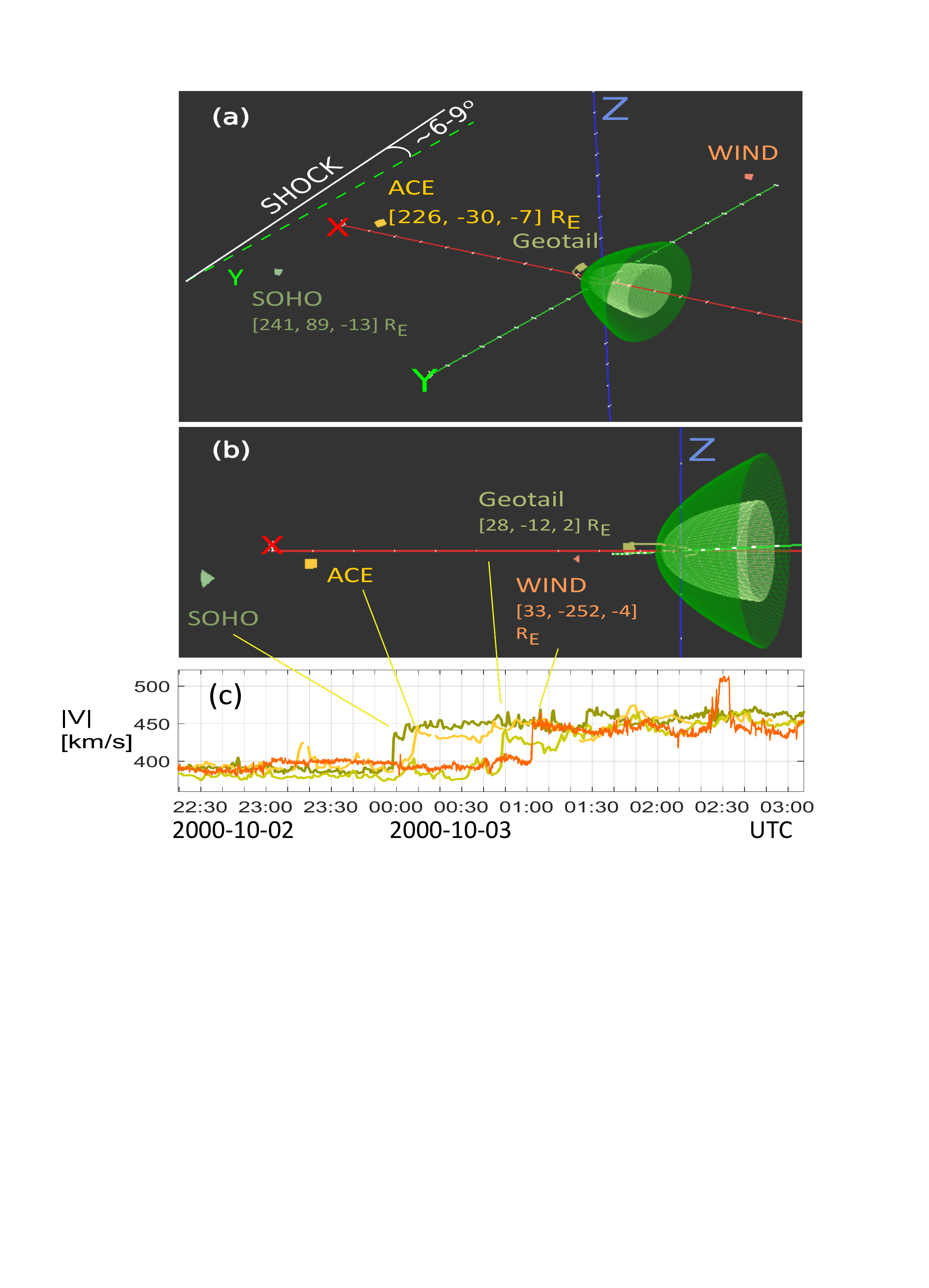}
 \caption{(a - b) Spacecraft locations in GSE (Geocentric Solar Ecliptic) coordinates; The shock front is deviated from the -Y direction by 6$^{\circ}$-9$^{\circ}$ in counter-clockwise direction. (c) Observations of the ICME shock (jump in the magnitude of bulk velocity) by different spacecraft.
 }
 \label{fig1}
  \end{figure}

 \begin{figure}[h]
 \includegraphics[width=15cm]{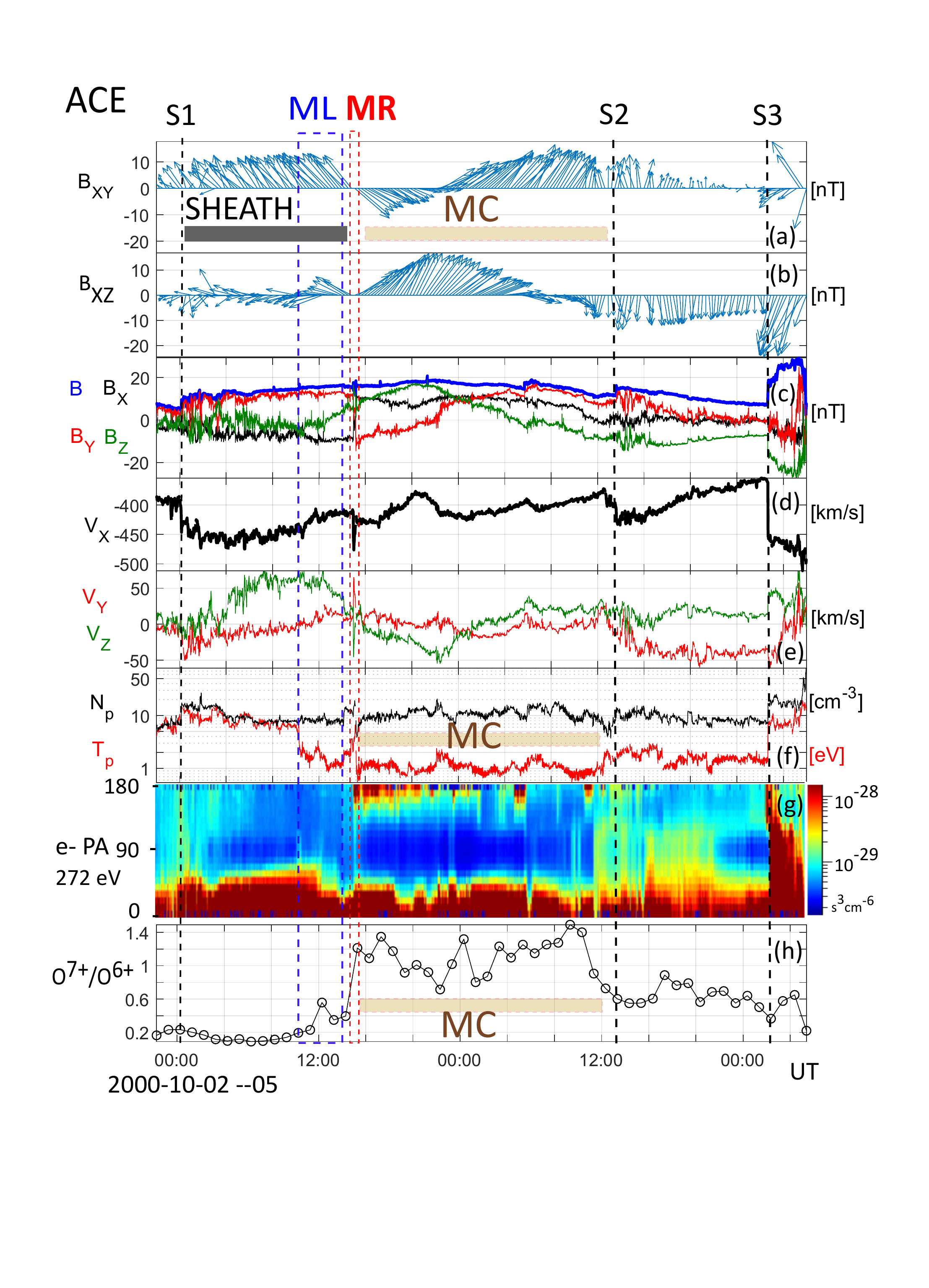}
 \caption{Event overview - ACE measurements. (a) B$_{XY}$ magnetic feather plot; (b) B$_{XZ}$ magnetic feather plot; (c) magnetic field magnitude $\mathbf{|B|}$ and GSE $B_X$, $B_Y$ and $B_Z$ components; (d) proton velocity GSE $V_X$ component; (e) proton velocity GSE $V_Y$ and $V_Z$ components; (f) proton density N$_p$ and temperature T$_p$ (the same Y axis); (g) e-PA: pitch-angle distribution of 272 eV electrons; (h) Oxygen ion charge-state ratio $O^{7+}/O^{6+}$; S1 - MC driven fast forward shock; S2 and S3 - fast forward shocks driven by other ejecta; ML - mixing layer; MR - magnetic reconnection; MC - magnetic cloud.
 }
 \label{fig2}
  \end{figure}

 \begin{figure}[h]
 \includegraphics[width=14cm]{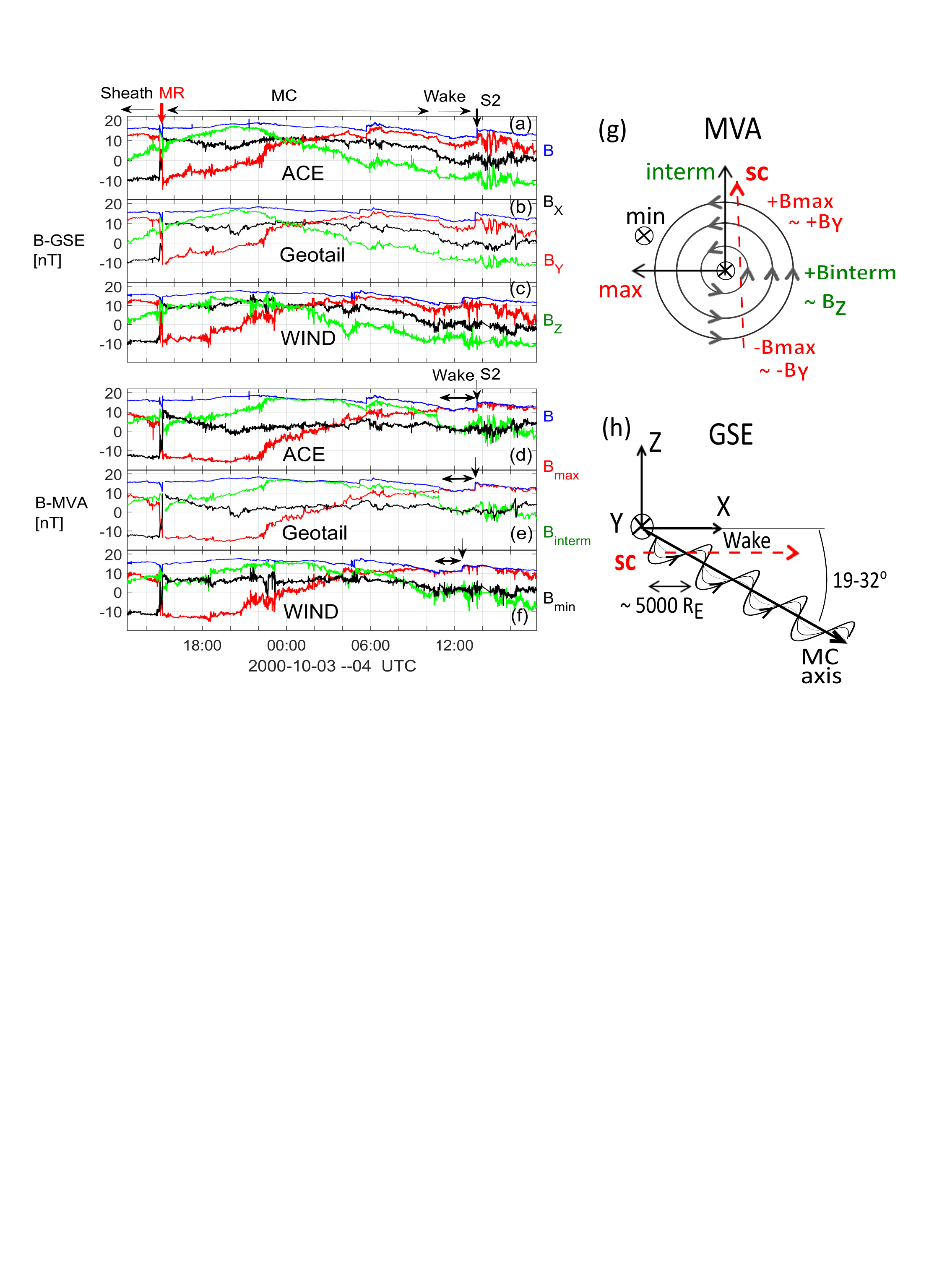}
 \caption{MC observations by ACE, Geotail and WIND spacecraft. Geotail and WIND data are time-shifted to match ACE observation of MR (MC boundary). (a - c) magnetic field magnitude B and GSE $B_X$, $B_Y$ and $B_Z$ components; (d - e) magnetic field magnitude $\mathbf{|B|}$ and MVA maximum, intermediate and minimum magnetic field components, $B_{max}$, $B_{interm}$ and $B_{min}$, respectively. Between the shock S2 and the end of MC a wake region can be observed. (g) cross-section of the MC in maximum and intermediate MVA coordinates; the dashed red arrow shows the crossing of the spacecraft across the structure; (h) the helical magnetic field and the direction of MC axis in GSE X-Z plane; half-turning of the helical magnetic field is $\sim$5000$R_E$; spacecraft crossing and wake location is also indicated.
 }
 \label{fig3}
  \end{figure}

 \begin{figure}[h]
 \includegraphics[width=15cm]{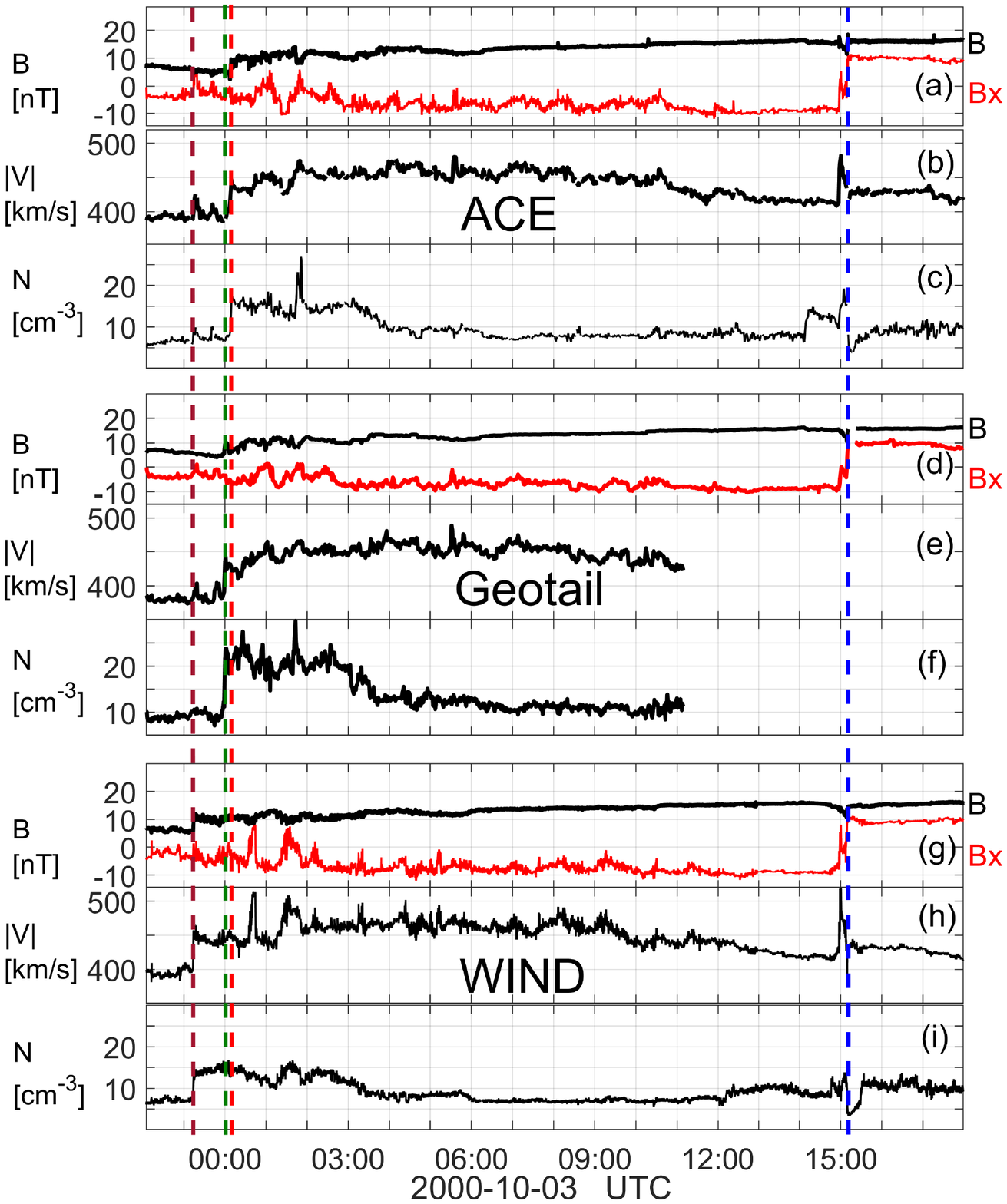}
 \caption{Estimation of the ICME sheath width along trajectories of ACE, Geotail and WIND. Geotail and WIND data are shifted to match ACE observation of MR (MC boundary) - blue vertical dashed line.
 Magnetic field magnitude, GSE B$_X$, magnitude of bulk velocity and density are shown for (a-c) ACE; (d-f) Geotail; (g-i) WIND. The MC driven shock observations by spacecraft are indicated by vertical dashed lines.  The sheath is wider (in time)  along WIND crossing than along ACE or Geotail crossings.
 }
 \label{fig4}
  \end{figure}

   \begin{figure}[h]
 \includegraphics[width=15cm]{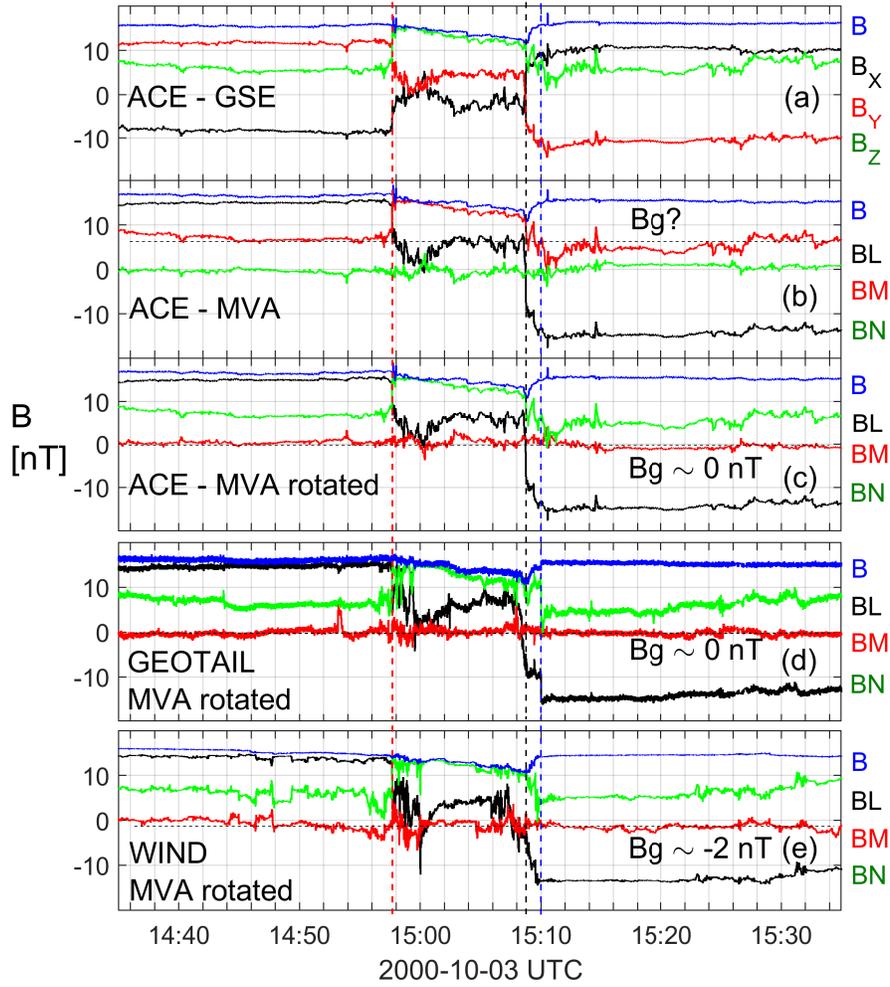}
 \caption{MR bifurcated current sheet. Geotail and WIND data are time shifted to match ACE observations at red dashed vertical line (a) ACE magnetic field magnitude B and GSE $B_X$, $B_Y$ and $B_Z$ components; (b) ACE magnetic field magnitude B and MVA $B_L$, $B_M$ and $B_N$ components;
 (c-e) ACE, Geotail and WIND magnetic field magnitude B and  $B_L$, $B_M$ and $B_N$ components in MVA rotated system; Bg is the guide field; the black and blue vertical dashed lines correspond to the right boundary of the current sheet as observed by different spacecraft.
 }
 \label{fig5}
  \end{figure}

      \begin{figure}[h]
 \includegraphics[width=15cm]{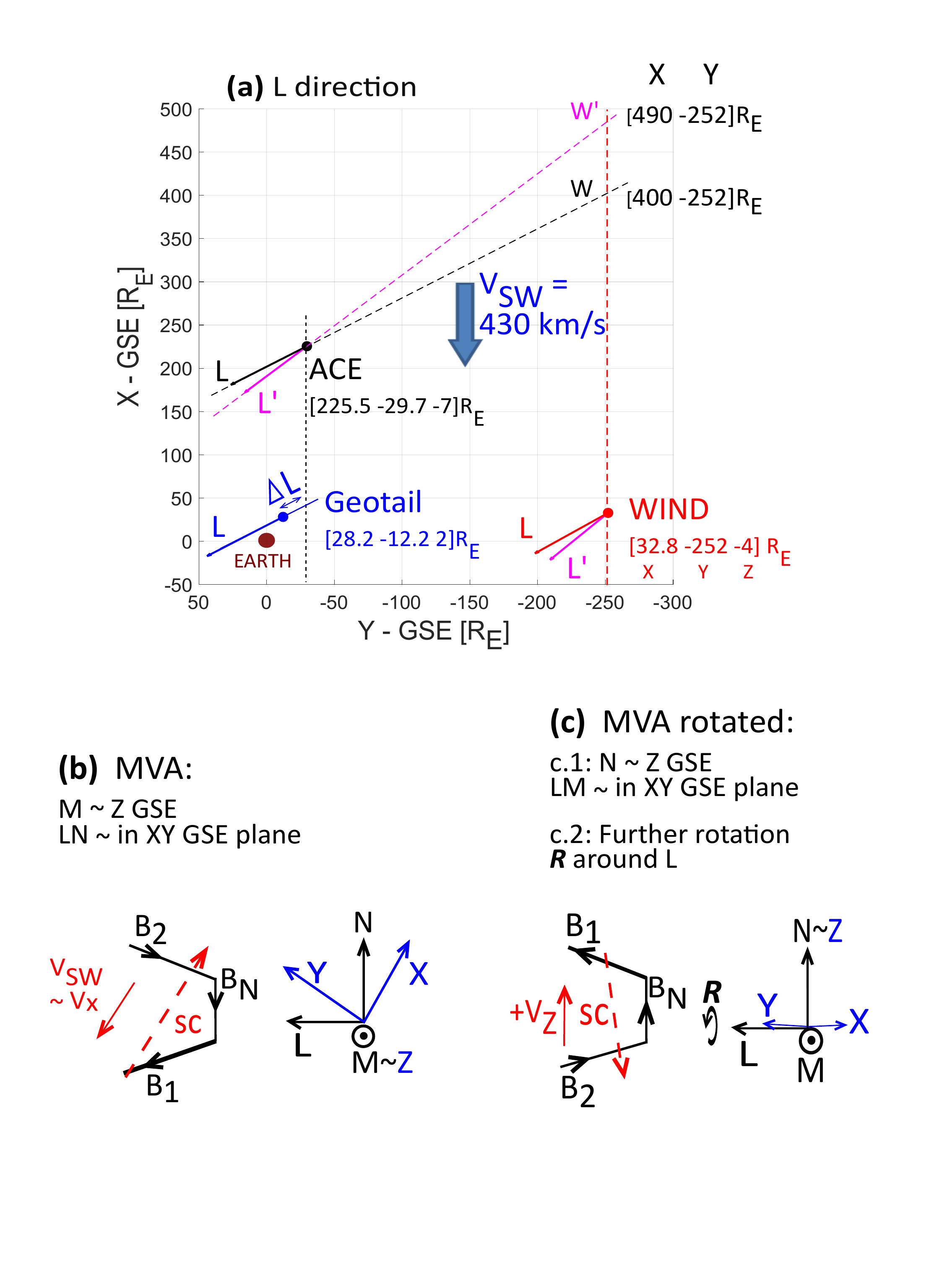}
 \caption{MVA direction alternatives. (a) Spacecraft positions in Y-X GSE plane and the local MVA $\mathbf{L}$  directions. $\mathbf{L'}$  directions  are estimated from $\mathbf{L}$  directional perpendicular outflow speeds from ACE and WIND plasma data. The W and W' points  are determined as projections of ACE $\mathbf{L}$  and $\mathbf{L'}$  directions to  upstream Y-GSE of WIND. (b) Current sheet in MVA coordinates; here the LN plane lies approximately in XY-GSE plane; the structure is convected by the bulk solar wind velocity. (c) Current sheet in MVA rotated coordinates; Two cases are shown: (c.1) here the LM plane lies approximately in XY-GSE plane; the structure is convected vertically; (c.2) Further optional rotation of $\mathbf{M N}$  around $\mathbf{L}$  is allowed.
 }
 \label{fig6}
  \end{figure}

     \begin{figure}[h]
 \includegraphics[width=15cm]{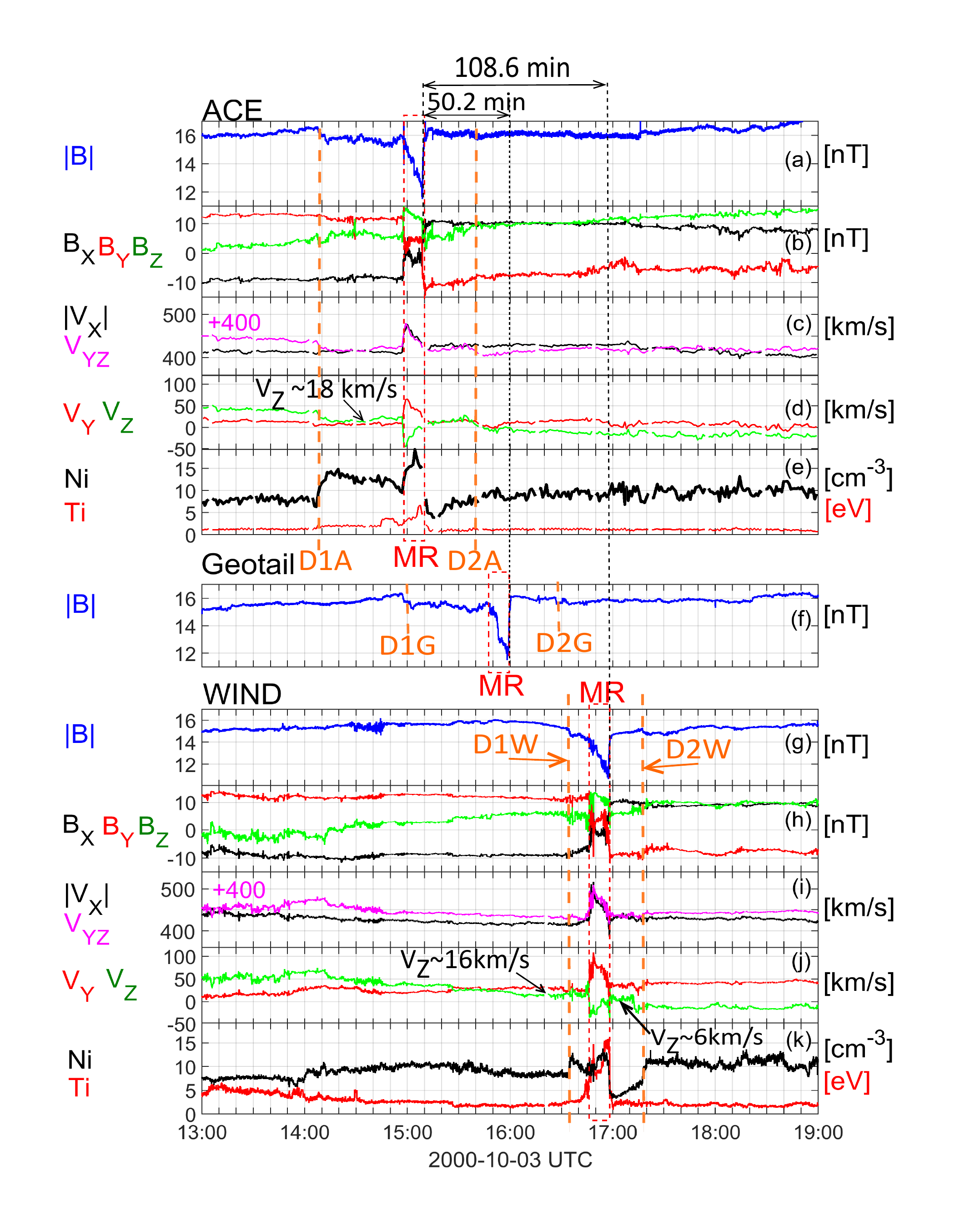}
 \caption{The vicinity of MR outflow observed by ACE, Geotail and WIND in GSE. (a-b) ACE magnetic field observations; (c) bulk velocity $|V_X|$ and $V_{YZ}=sqrt(V_Y^2+V_Z^2)$; (d) $V_Y$ and $V_Z$; (e) ion (proton) density and temperature, Ni and Ti (the same Y axis is used); (f) magnitude of the magnetic field from Geotail; (g-k) WIND observations of the same quantities as above for ACE.   The time difference between MR observations of ACE-Geotail is 50.2 minutes, ACE-WIND is 108.6 minutes. Earlier in time than MR the spacecraft are in the ICME sheath; Later in time than MR the spacecraft are in the MC side. Discontinuities in the sheath are marked as D1A, D1G and D1W, in the MC are marked as D2A, D2G and D2W, respectively (see Tables 2 and 3).
 }
 \label{fig7}
  \end{figure}

   \begin{figure}[h]
 \includegraphics[width=14cm]{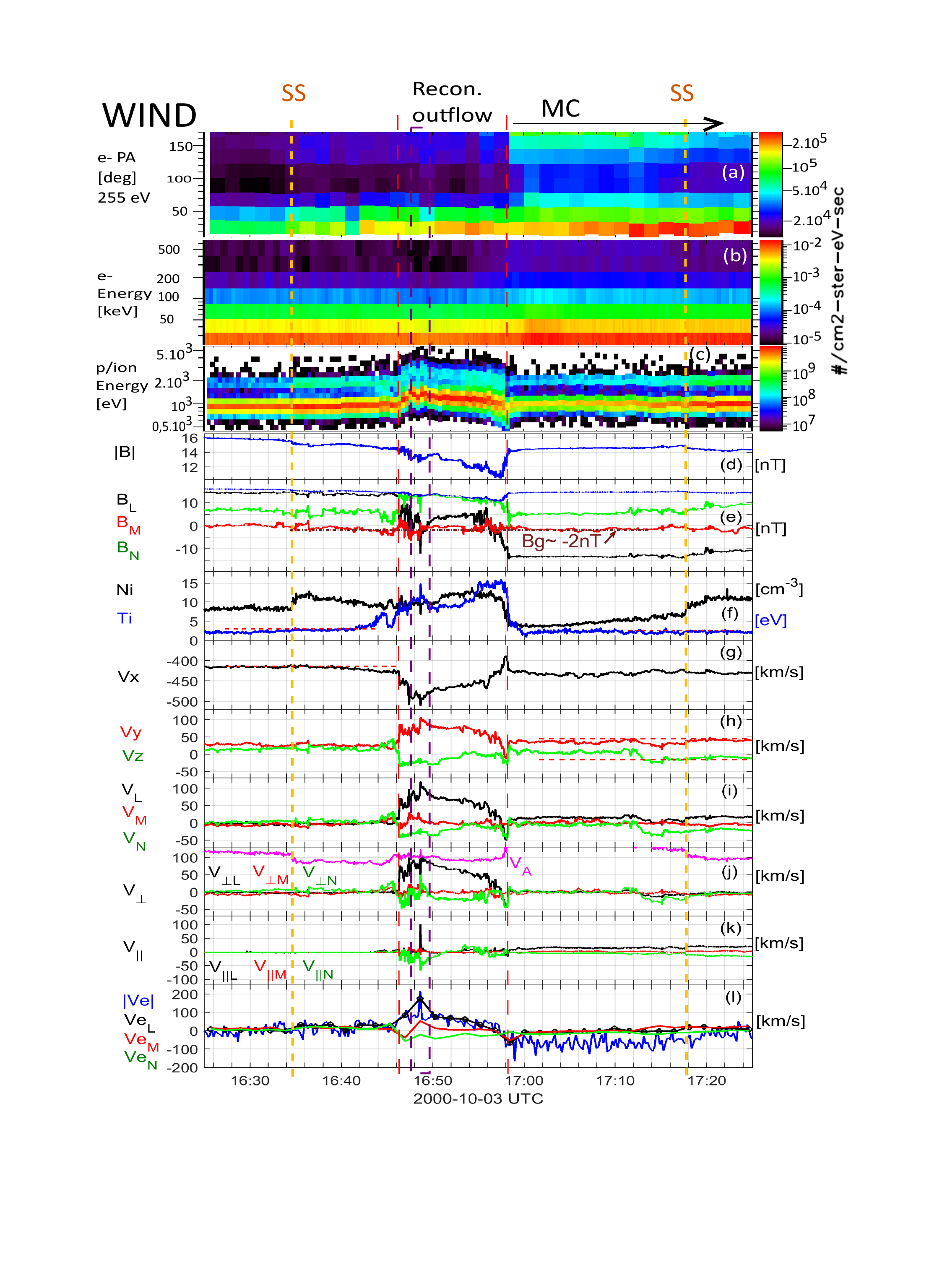}
 \caption{WIND observations of MR outflow: particle, field and flow signatures. (a) Pitch-angle (e-PA) distribution of 255 eV electrons; (b) omni-directional electron energy spectrum; (c) omni-directional proton/ion energy spectrum; (d) magnitude of the magnetic field; (e) MVA rotated LMN components of the magnetic field; (f) ion density and temperature with the same Y axis; (g-h) GSE XYZ components of the ion bulk velocity; (i) MVA rotated  LMN components of the ion velocity, ambient velocity averages removed; (j) perpendicular to magnetic field  LMN velocities and the local Alfv\'en velocity $V_A$; (k) parallel to magnetic field  LMN velocities; (l) on-board computed magnitude of the electron velocity $|Ve|$ and the ground calculated electron velocity LMN components. The dashed vertical yellow lines indicate the SS-like discontinuities (D1W and D2W). The dashed vertical red lines show the outflow boundaries. The dashed magenta box contains the electron jet and accelerated protons.
 }
 \label{fig8}
  \end{figure}

\begin{figure}[h]
 \includegraphics[width=14cm]{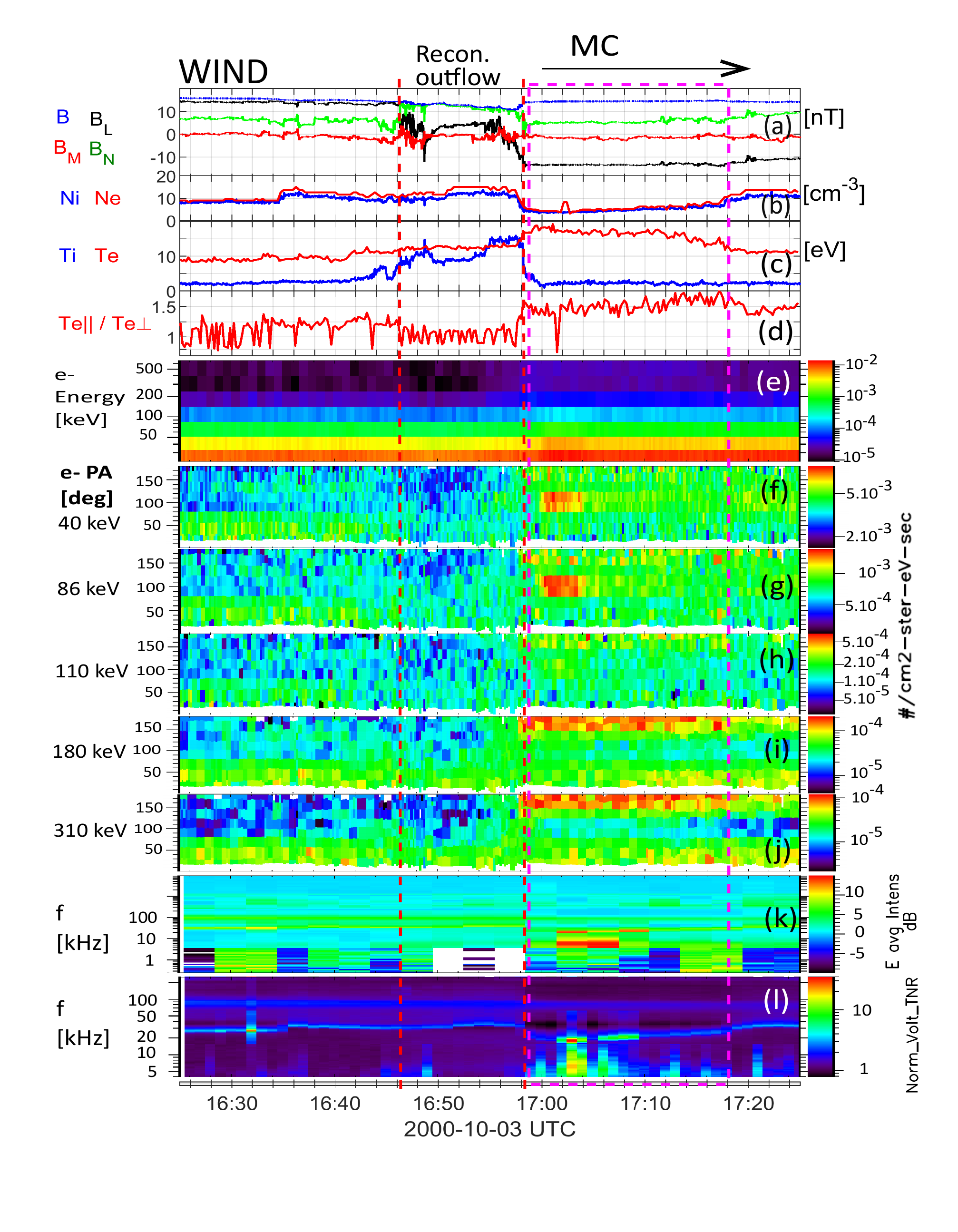}
 \caption{WIND observations of MR outflow: physical processes near the MR separatrix/inflow region in the MC boundary layer. (a)  magnitude and the MVA rotated LMN components of the magnetic field; (b) ion and electron densities, the latter estimated from WIND/WAVE experiment; (c) ion and electron temperatures; (d) electron parallel to perpendicular temperature ratio (anisotropy); (e) omni-directional electron energy spectrum; (f-j) e-PA electron pitch-angle distributions for
 energies 40, 86, 110, 180, 310 keV; (k) spectrum of the average electric field fluctuations    above the background in dB; (l) the normalized average voltage from the Thermal Noise Receiver  experiment on Wind/WAVES. The dashed vertical red lines show the outflow boundaries. The dashed magenta box shows the density cavity.
  }
 \label{fig9}
  \end{figure}

\begin{figure}[h]
 \includegraphics[width=14cm]{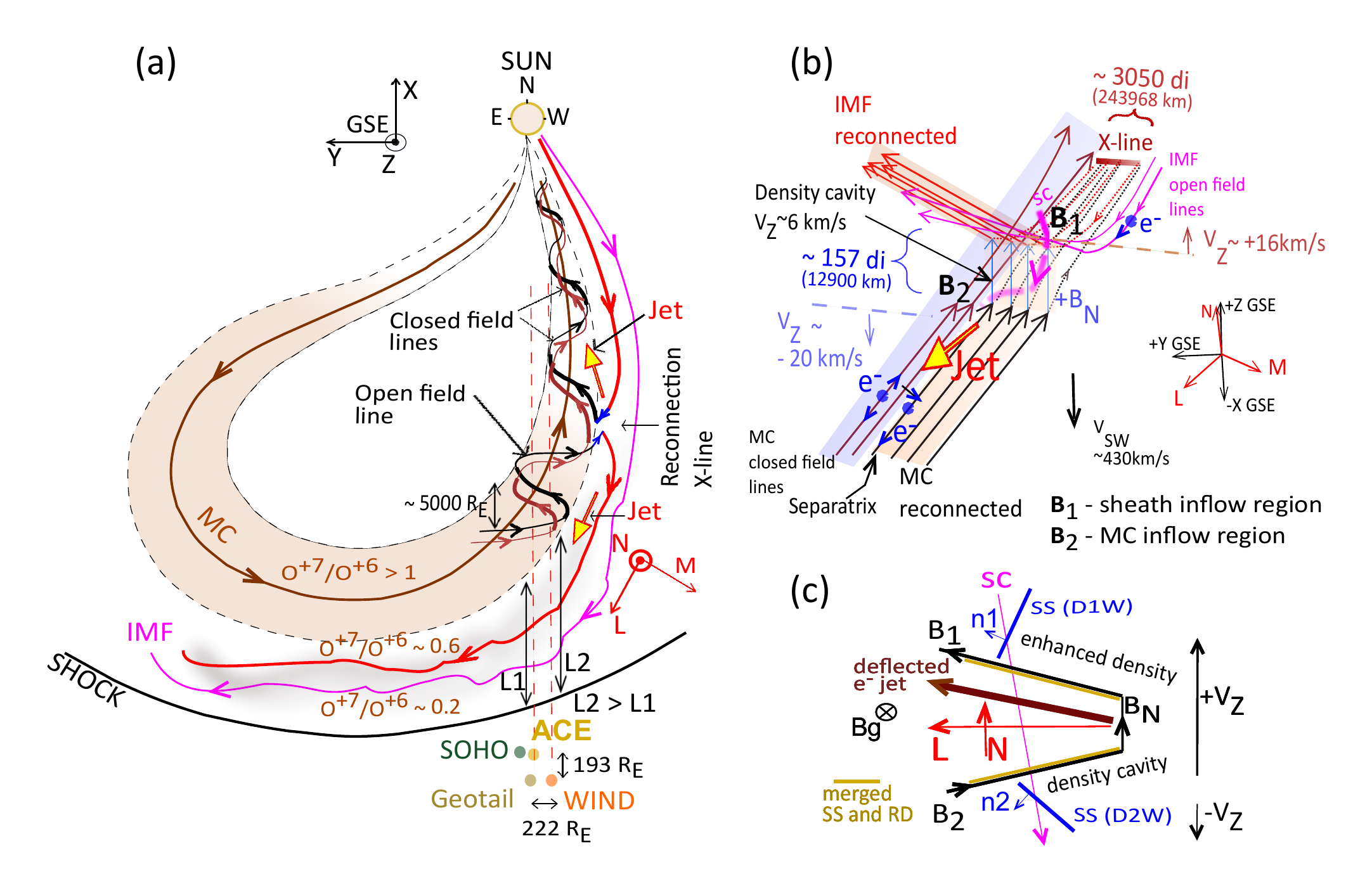}
 \caption{Event summary cartoons. (a) The large-scale structure of ICME and the left-handed helical MC with axis pointing towards Sun. Magnetic field lines over the MC axis are thicker, below thinner. MR occurs at the western flank of the MC. The relative positions of SOHO, ACE, Geotail and WIND spacecraft are shown on the bottom. The reconnection X-line distance from spacecraft crossings of the structure is unknown, but much larger than 314$R_E$. For simplicity two trajectories (ACE and WIND)  are shown only (red vertical dashed lines). Geotail and SOHO trajectories are close to ACE trajectory. The sunward jet (outflow) is associated with closed, the antisunward jet with open field lines. On open freshly reconnected field lines the leakage of Oxygen ions occurs. The sheath is wider along WIND trajectory (sheath width: L2) than along ACE/Geotail trajectories (sheath width: L1);  (b) Reconnecting current sheet 3D geometry (here the spatial segment includes the X-line and the outflow crossing from ICME sheath to MC); spacecraft trajectory (sc-magenta line, it is indicating a crossing by a spacecraft) is from $\mathbf{B_1}$ to $\mathbf{B_2}$;  current sheet thickness is $\sim$157 di; X-line length is $\sim$3050 di; parallel, anti-parallel, perpendicular electron populations are shown on MC closed field lines, inflow and separatrix regions and on open IMF field lines; $V_Z$ changes across the structure and the location of density cavity near $\mathbf{B_2}$ are indicated. (c) 2D cartoon of reconnecting current sheet in LN plane; the subplot shows the Lorentz force deflected electron jet, the SS-like discontinuities, the merged SS and RD, the regions of density enhancement and cavity and the $\pm$ $V_Z$.
 }
 \label{fig10}
  \end{figure}

\acknowledgments
Z.V. and Y.N. were supported by the Austrian FWF under contract P28764-N27. E. Y. was supported by the Swedish Civil Contingencies Agency, grant 2016-2102. The authors would like to thank Teimuraz Zaqarashvili (University of Graz) for the helpful suggestions regarding KH instability. ACE, WIND field and plasma data used in this paper are  available from https://cdaweb.gsfc.nasa.gov/. ACE/SWEPAM data are available from http://www.srl.caltech.edu/ACE/ASC/DATA/level3/swepam/. Geotail data are available from https://darts.isas.jaxa.jp/stp/geotail/. SOHO high resolution plasma data are available from https://l1.umd.edu/.


%
%




\end{document}